\documentclass[preprint,aps,showpacs]{revtex4}
\usepackage{epsfig}
\usepackage{bm}

\begin{document}

\title{
Proton polarizations in polarized \bm{$^3$}He
studied with the 
\bm{$ \vec{^3{\rm He}} ({\vec e},e' p)d$}
and
\bm{$ \vec{^3{\rm He}} ({\vec e},e' p)pn$}
processes
}

\author{J.~Golak, R.~Skibi\'nski, H.~Wita{\l}a}
\affiliation{M. Smoluchowski Institute of Physics, Jagiellonian University,
                    PL-30059 Krak\'ow, Poland}

\author{W.~Gl\"ockle}
\affiliation{Institut f\"ur Theoretische Physik II,
         Ruhr Universit\"at Bochum, D-44780 Bochum, Germany}

\author{A.~Nogga}
\affiliation{Forschungszentrum J\"ulich, IKP
(Theorie), D-52425 J\"ulich, Germany}

\author{H.~Kamada}
\affiliation{Department of Physics, Faculty of Engineering,
  Kyushu Institute of Technology,
  1-1 Sensuicho, Tobata, Kitakyushu 804-8550, Japan}

\date{\today}

\begin{abstract}
We study 
within the Faddeev framework the 
$ \vec{^3{\rm He}} ({\vec e},e' p)d$
as well as the
$ \vec{^3{\rm He}} ({\vec e},e' p)pn$
and
$ \vec{^3{\rm He}} ({\vec e},e' n)pp$
reactions in order to extract information on the proton
and neutron polarization in polarized $^3$He. 
We achieve clear analytical insight for simplified dynamical assumptions
and define conditions for experimental access to important 
$^3$He properties. 
In addition we point to the possibility to measure the
electromagnetic proton form factors in the process 
$ ^3{\rm He} (e,e' p)d$
which would test the dynamical picture and put limits on
medium corrections of the form factors.
\end{abstract}
\maketitle



\section{Introduction}
\label{sec:1}

With the possibility of solving precisely few-nucleon equations and
the availability of high precision nucleon-nucleon potentials 
it is tempting to ask very detailed questions about the properties
of light nuclei.  
Spin dependent momentum distributions
of nuclear clusters inside light nuclei 
have been studied at many places, see for instance \cite{Fentometer}. 
Especially the $^3$He nucleus is interesting. The availability of
highly polarized $^3$He allows one to perform very detailed electron 
scattering experiments, which, due to the recent progress in the 
calculations of three-nucleon (3N) bound and scattering
states, can be analyzed very precisely. This makes it tempting 
to extract information on its properties.

In a recent paper \cite{spindep} we addressed the question whether 
momentum distributions of polarized proton-deuteron (pd) clusters 
in polarized $^3$He could be accessed through 
the $ \vec{^3{\rm He}} (e,e' {\vec p})d$
or $ \vec{^3{\rm He}} (e,e' {\vec d})p$
processes. Final state interactions (FSI) and meson exchange currents (MEC) 
turned out to destroy the clear picture offered by the plane wave impulse approximation (PWIA)
and the assumption of the single nucleon current operator. This we found 
for most of the cases studied in the kinematical regime below the pion production
threshold. Only for small relative pd momenta 
direct access to the sought $^3$He properties appeared possible. 
The 
$ \vec{^3{\rm He}} (e,e' {\vec p})d$
or $ \vec{^3{\rm He}} (e,e' {\vec d})p$
experiments 
would require, however, measuring the polarizations of the outgoing
particles, which is very demanding.

In this paper we would like to investigate theoretically two processes,
$ \vec{^3{\rm He}} ({\vec e},e' p)d$
and $ \vec{^3{\rm He}} ({\vec e},e' p)pn$,
measured recently at MAMI~\cite{Achenbach05}
and show that under the same PWIA assumption they
provide equivalent information about $^3$He properties.
We remind the reader of our formalism in Sec.~\ref{sec:2}.
Section~\ref{sec:3} shows our results for the exclusive proton-deuteron 
breakup of $^3$He and Sec.~\ref{sec:4} deals with different aspects 
of the semi-exclusive $ \vec{^3{\rm He}} ({\vec e},e' p)pn$ 
reaction.  We end with a brief summary in Sec.~\ref{sec:5}.

\section{Theory}
\label{sec:2}

The spin dependent momentum distributions of proton-deuteron
clusters inside the $^3$He nucleus are defined as:
\begin{equation}
{\cal Y} ( m_3, m_d, m_p ; {\vec q} \, ) \ \equiv \
\left\langle \Psi m_3 \left|
| \phi_d m_d  \rangle 
| {\vec q} \, \frac12 m_p  \rangle  
\langle {\vec q} \,  \frac12 m_p |
\langle \phi_d m_d |
\right| \Psi m_3 \right\rangle ,
\label{eq1}
\end{equation}
where $ {\vec q} $ is the proton momentum
(the deuteron momentum is $-{\vec q} $); $m_p$, $m_d$ and $m_3$ are
spin magnetic quantum numbers for the proton, deuteron
and $^3$He, respectively.
These quantities can be written as 
\begin{eqnarray}
{\cal Y} ( m_3, m_d, m_p ; {\vec q} \, ) \, = \,
\left| 
\sum_{\lambda = 0,2 } \, 
Y_{ \lambda , m_3 - m_d - m_p } ( {\hat q} ) \,
C( 1 I_\lambda \frac12 ; m_d , m_3 - m_d , m_3 ) \right. \cr
\left. C( \lambda \frac12 I_\lambda ; m_3 - m_d - m_p , m_p, m_3 - m_d ) 
\ H_\lambda (q) \right|^2 ,
\label{eq3.3}
\end{eqnarray}
where $H_\lambda (q) $ is the overlap of the deuteron state
and the $^3$He state calculated in momentum space~\cite{Gloecklebook}
\begin{eqnarray}
H_\lambda (q) \,\equiv \, 
\sum_{l = 0,2 } \,
\int_0^\infty d p \, p^2 \,  \phi_l (p) \,
\langle p q \alpha_{l \lambda} | \Psi \rangle \ \ , \ \lambda = 0, 2.
\label{eq3.4}
\end{eqnarray}
Here $ \langle p q \alpha | \Psi \rangle $ are the partial
wave projected wave function components of $^3$He and $  \phi_l (p) $
are the s- and d-wave components of the deuteron.
The set $ \alpha_{l \lambda} $ contributes only for the deuteron
quantum numbers $s=1$, $j=1$ and $t=0$. 
Further $I_\lambda = \frac12 $ for $\lambda =0$ and $\frac32 $ for $\lambda =2$.
It is clear that using this quantity $H_\lambda (q) $
the spin dependent momentum distribution
$ {\cal Y} ( m_3, m_d, m_p ; {\vec q}\,  ) $
can be constructed for any combination of magnetic
quantum numbers and any direction ${\hat q}$.

In \cite{spindep} we also showed that under the 
PWIA treatment and in the non-relativistic limit 
there are simple relations between different $ {\cal Y} $'s and
the response functions $W_i$, 
which enter the laboratory cross section for the process
$ {\vec e} + \vec{{}^3{\rm He}} \rightarrow e' + p + d$. 
This cross section has the form~\cite{Donnelly}
\begin{eqnarray}
\sigma ({\vec S} , h) \ & = & \ 
\sigma_{\rm Mott} \, 
\left\{ \,
\left( v_L W_L + v_T W_T + v_{TT} W_{TT} + v_{TL} W_{TL} \right) \, \right. \nonumber \\
& + & \left. h \, \left( v_{T'} W_{T'} + v_{TL'} W_{TL'} \right) \,
\right\} \,
\rho ,
\label{eq4}
\end{eqnarray}
where $\sigma_{\rm Mott}$, $v_i$ and $\rho$ are analytically given kinematical factors,
$h$ is the helicity of the incoming electron
and ${\vec S}$ represents the initial $^3$He spin direction.

This means that both the cross section 
and the helicity asymmetry $ A ( {\vec S} ) $ 
\begin{eqnarray}
A ( {\vec S} ) \equiv 
\frac{\sigma ({\vec S} , h=+1) - \sigma ({\vec S} , h=-1)}{ \sigma ({\vec S} , h=+1) + \sigma ({\vec S} , h=-1) }
\label{A}
\end{eqnarray}
for the $ \vec{^3{\rm He}} ({\vec e},e' p)d$ process
can be obtained, assuming PWIA, in terms of $ H_\lambda $,
the electromagnetic proton form factors $G_E^p$ and $G_M^p$,
and simple kinematical quantities. The response functions read
\begin{eqnarray}
W_L =
\frac{{\,{G_E^p}}^2\,\left( {\,{H_0 (q)}}^2 + 
      {\,{H_2 (q)}}^2 \right) }{4\,\pi }
\label{wl}
\end{eqnarray}
\begin{eqnarray}
W_T =
\frac{\left( {\,{H_0 (q)}}^2 + {\,{H_2 (q)}}^2 \right) \,
      \left( 
        {\,{G_M^p}}^2\,{\,{Q}}^2  + 
        {\,{G_E^p}}^2\,{\,{q_f}}^2 - 
        {\,{G_E^p}}^2\,{\,{q_f}}^2\,\cos (2\,\,{\theta_1})
  \right) }{8\,{\,{M}}^2\,\pi }
\label{wt}
\end{eqnarray}
\begin{eqnarray}
W_{TT} =
\frac{- {\,{G_E^p}}^2\,
      \left( {\,{H_0 (q)}}^2 + {\,{H_2 (q)}}^2 \right) \,
      {\,{q_f}}^2\,\cos (2\,\phi)\,{\sin^2  (\,{\theta_1})} \ }
        {4\,{\,{M}}^2\,\pi }
\label{wtt}
\end{eqnarray}
\begin{eqnarray}
W_{TL} =
\frac{{\,{G_E^p}}^2\,\left( {\,{H_0 (q)}}^2 + 
      {\,{H_2 (q)}}^2 \right) \,\,{q_f}\,\cos (\phi)\,
    \sin (\,{\theta_1})}{{\sqrt{2}}\,\,{M}\,\pi }
\label{wtl}
\end{eqnarray}
\begin{eqnarray}
W_{T^\prime} = \frac{B_1 \, \cos \theta^\star \ + 
\ B_2 \, \sin  \theta^\star \cos ( \phi - \phi^\star )  }
{48 \, \pi \, M^2}
\label{wtp}
\end{eqnarray}
\begin{eqnarray}
W_{{TL}^\prime} = 
\frac{
{C}_1 \, \cos ( 2 \phi - \phi^\star ) \, \sin (\theta^\star ) \, + \,
{C}_2 \, \cos ( \phi^\star ) \, \sin (\theta^\star ) 
}
{48 \, \pi \, M} \nonumber \\
+ \
\frac{
{C}_3 \, \cos (\theta^\star ) \, + \,
{C}_4 \, \cos ( \phi - \phi^\star ) \, \sin (\theta^\star ) 
}
{48 \, \pi \, M} .
\label{wtplp}
\end{eqnarray}
The auxiliary quantities $B_1$, $B_2$, $C_1$--$C_4$, 
which appear in the helicity-dependent response functions 
$W_{{T}^\prime}$ and $W_{{TL}^\prime}$
in Eqs.~(\ref{wtp}) 
and (\ref{wtplp}) are
\begin{eqnarray}
B_1 =
\, \left( {G_M^p}\, Q \right)^2 \, \left( 2\,{\,{H_0 (q)}}^2 +
     2\,{\sqrt{2}}\,\,{H_0 (q)}\,\,{H_2 (q)} + {\,{H_2 (q)}}^2 \
\right) \, \nonumber \\
+ \ 3\,
   \left( 2\,{\sqrt{2}}\,\,{H_0 (q)} - \,{H_2 (q)} \right) \,
   \,{H_2 (q)}\, \left( {G_M^p}\,{Q} \right)^2 \, \cos (2\,\theta) 
\nonumber \\
- \ 6\,
   \left( 2\,{\sqrt{2}}\,\,{H_0 (q)} - \,{H_2 (q)} \right) \,
   \,{H_2 (q)}\, {G_E^p}\,\,{q}\, {G_M^p}\,{Q} \,
      \sin (2\,\theta)\,\sin (\,{\theta_1}) 
\label{b1}
\end{eqnarray}
\begin{eqnarray}
B_2 =
\ -3\,\, \left( {G_M^p}\, Q \right)^2 \,{H_2 (q)}\,
   \left( -2\,{\sqrt{2}}\,\,{H_0 (q)} + \,{H_2 (q)} \right) \,
   \,\sin (2\,\theta)  \nonumber \\
- \  2\,\,{G_E^p}\,\,{q_f}\, {G_M^p}\, Q \,
   \left( 2\,{\,{H_0 (q_f)}}^2 + 
     2\,{\sqrt{2}}\,\,{H_0 (q_f)}\,\,{H_2 (q_f)} + 
   {\,{H_2 (q_f)}}^2 \right) \, \sin (\,{\theta_1})
          \nonumber \\
- \ 6\,\,{G_E^p}\,\,{q_f}\, {G_M^p}\, Q \,
     \,{H_2 (q)}\,\left( -2\,{\sqrt{2}}\,\,{H_0 (q)} + \,{H_2 (q)} \right) \,
      \cos (2\,\theta) 
   \sin (\,{\theta_1})
\end{eqnarray}
\begin{eqnarray}
{C}_1 = 
3\,\,{G_E^p}\,\,{G_M^p}\,\,{H_2 (q)}\,
  \left( -4\,\,{H_0 (q)} + {\sqrt{2}}\,\,{H_2 (q)} \right) \,
  \,{Q}
\end{eqnarray}
\begin{eqnarray}
{C}_2 = 
\,{G_E^p}\,\,{G_M^p}\,
  \left( 4\,\,{H_0 (q)}\,\,{H_2 (q)} - 
    {\sqrt{2}}\,\left( 4\,{\,{H_0 (q)}}^2 + 5\,{\,{H_2 (q)}}^2 \
\right)  \right) \,\,{Q}
\end{eqnarray}
\begin{eqnarray}
{C}_3 = 
6\,\,{G_E^p}\,\,{G_M^p}\,\,{H_2 (q)}\,
  \left( -4\,\,{H_0 (q)} + {\sqrt{2}}\,\,{H_2 (q)} \right) \,
  \,{Q}\,\cos (\phi)\,\sin (2\,\theta)
\end{eqnarray}
\begin{eqnarray}
{C}_4 = 
-6\,\,{G_E^p}\,\,{G_M^p}\,\,{H_2 (q)}\,
  \left( -4\,\,{H_0 (q)} + {\sqrt{2}}\,\,{H_2 (q)} \right) \,
  \,{Q}\,\cos (\phi)\,\cos (2\,\theta) .
\label{c4}
\end{eqnarray}
We assume the reference frame, for which the
three-momentum transfer $\vec{Q} \equiv {\vec k} - {\vec k}^{\, \prime} $  
is parallel to $\hat{z}$
and $\hat{y} \equiv \frac{ {\vec k}^{\, \prime} \times {\vec k}}
   {\mid {\vec k}^{\, \prime} \times {\vec k} \mid}$,
as well as  $\hat{x} = \hat{y} \times \hat{z}$.
Here $ {\vec k}$ and ${\vec k}^{\, \prime} $ are the initial and final
electron momenta.
In this system  
$\theta_1$ and $\phi_1$ are the polar and azimuthal angles 
corresponding to the direction of the final proton-deuteron relative momentum
$\vec{q}_f  \equiv \vec{p}_p - \frac13 \vec{Q} $, 
$\theta$ and $\phi$ are the polar and azimuthal angles 
corresponding to the direction of
$ \vec{q} \equiv \vec{q}_f - \frac23  \vec{Q} = \vec{p}_p - \vec{Q} = - \vec{p}_d$. 
The initial $^3$He spin orientation is defined in terms 
of the $\theta^\star$ and $\phi^\star$ angles.
Further, 
$Q \equiv \mid \vec{Q} \mid$,
$ q_f \equiv \mid {\vec q}_f \mid$ 
and 
$q \equiv \mid  \vec{q}_f - \frac23  \vec{Q} \mid = \mid \vec{p}_d \mid $,
where $  \vec{p}_p $ and $  \vec{p}_d $ are the final proton 
and deuteron momenta. Finally $M$ is the nucleon mass.

These expressions simplify significantly if the so-called parallel kinematics
is assumed, for which the final proton is ejected parallel to $ \vec{Q} $.
Then $\theta = \theta_1 = \phi = \phi_1 = 0$ and
the response functions given in Eqs.~(\ref{wt})--(\ref{wtplp}) 
and (\ref{b1})--(\ref{c4})
reduce to

\begin{eqnarray}
W_T =
\frac{ \left(  \,G_M^p \,Q \right)^2 \,
         \left( {\,{H_0 (q)}}^2 + {\,{H_2 (q)}}^2 \right) \,
 }{8\,{\,{M}}^2\,\pi }
\label{wt.2}
\end{eqnarray}
\begin{eqnarray}
W_{TT} = 0
\label{wtt.2}
\end{eqnarray}
\begin{eqnarray}
W_{TL} = 0
\label{wtl.2}
\end{eqnarray}
\begin{eqnarray}
W_{T^\prime} = 
\frac{
\left( {G_M^p}\, Q \right)^2 \, \left( 2\,{\,{H_0 (q)}}^2 +
     2\,{\sqrt{2}}\,\,{H_0 (q)}\,\,{H_2 (q)} + {\,{H_2 (q)}}^2 \
\right) \, \, \cos ( \theta^\star ) } 
{48 \, \pi \, M^2}
\nonumber \\
+ \ \frac{
 3\,  \left( {G_M^p}\,{Q} \right)^2 \,
     \left( 2\,{\sqrt{2}}\,\,{H_0 (q)} - \,{H_2 (q)} \right) \,
   \,{H_2 (q)}\, 
 \, \cos ( \theta^\star ) } {48 \, \pi \, M^2} 
\label{wtp.2}
\end{eqnarray}
\begin{eqnarray}
W_{{TL}^\prime} = 
\frac{ 3\,\,{G_E^p}\,\,{G_M^p}\, Q \,{H_2 (q)}\,
  \left( -4\,\,{H_0 (q)} + {\sqrt{2}}\,\,{H_2 (q)} \right) \,
 \cos (\phi^\star ) \, \sin (\theta^\star )
}
{48 \, \pi \, M} \nonumber \\
+ \
\frac{
\,{G_E^p}\,\,{G_M^p}\, Q \,
  \left( 4\,\,{H_0 (q)}\,\,{H_2 (q)} - 
    {\sqrt{2}}\,\left( 4\,{\,{H_0 (q)}}^2 + 5\,{\,{H_2 (q)}}^2 \
\right)  \right) \,
\cos ( \phi^\star ) \, \sin (\theta^\star ) 
}
{48 \, \pi \, M} \nonumber \\
- \
\frac{
6\,\,{G_E^p}\,\,{G_M^p}\, Q \,{H_2 (q)}\,
  \left( -4\,\,{H_0 (q)} + {\sqrt{2}}\,\,{H_2 (q)} \right) \,
  \, \cos ( \phi^\star ) \, \sin (\theta^\star ) 
}
{48 \, \pi \, M} .
\label{wtplp.2}
\end{eqnarray}
This allows us to express the parallel and perpendicular 
helicity asymmetries 
in terms of $H_\lambda (q) $. For the parallel kinematics they are
\begin{eqnarray}
A_\parallel \equiv A ( \theta^\star = 0, \phi^\star = 0 ) =  & \nonumber \\
\frac{  \left( {G_M^p}\, Q \, \right)^2 \, 
\left( {H_0 (q)}^2 + 4\,{\sqrt{2}}\,{H_0 (q)}\,{H_2 (q)} - {H_2 (q)}^2 \, \right) \, v_{T^\prime}
}
{ 3 \, \left( {H_0 (q)}^2 + {H_2 (q)}^2 \right) \, 
\left( 2 \left( {G_E^p}\, \right)^2 \, M^2 \, v_L  + 
 \left( {G_M^p}\, Q \, \right)^2 \, v_T \right)
  } &
\label{apara}
\end{eqnarray}
\begin{eqnarray}
A_\perp \equiv A ( \theta^\star = \frac{\pi}{2}, \phi^\star = 0 ) =  & \nonumber \\
\frac{ -2 \, {G_E^p}\, {G_M^p}\, M \, Q \, 
\left( {\sqrt{2}}\, {H_0 (q)}^2 - 4\,{H_0 (q)}\,{H_2 (q)} + 2 \, {\sqrt{2}}\, {H_2 (q)}^2 \, 
\right) \, v_{{TL}^\prime}
}
{ 3 \, \left( {H_0 (q)}^2 + {H_2 (q)}^2 \right) \, 
\left( 2 \left( {G_E^p}\, \right)^2 \, M^2 \, v_L  + 
 \left( {G_M^p}\, Q \, \right)^2 \, v_T \right)
  } &
\label{aperp}
\end{eqnarray}
Here, the $^3$He wave function enters through the combinations
\begin{eqnarray}
P_1 \equiv 
\frac{ {H_0 (q)}^2 + 4\,{\sqrt{2}}\,{H_0 (q)}\,{H_2 (q)} - {H_2 (q)}^2 }
{ 3 \, \left( {H_0 (q)}^2 + {H_2 (q)}^2  \right)  } 
\label{p1}
\end{eqnarray}
and
\begin{eqnarray}
P_2 \equiv 
\frac{ {H_0 (q)}^2 - 2 \,{\sqrt{2}}\,{H_0 (q)}\,{H_2 (q)} + 2 \,{H_2 (q)}^2 }
{ 3 \, \left( {H_0 (q)}^2 + {H_2 (q)}^2  \right) }  .
\label{p2}
\end{eqnarray}
in terms of which 
\begin{eqnarray}
A_\parallel =
\frac{  
\left( {G_M^p}\, Q \, \right)^2 \, v_{T^\prime}
}
{ 
 2 \left( {G_E^p}\, \right)^2 \, M^2 \, v_L  + 
 \left( {G_M^p}\, Q \, \right)^2 \, v_T 
  } \ P_1
\label{Aparap1}
\end{eqnarray}
and
\begin{eqnarray}
A_\perp =
\frac{ 
-2 \, \sqrt{2} \, {G_E^p}\, {G_M^p}\, M \, Q \, v_{{TL}^\prime}
}
{ 
 2 \left( {G_E^p}\, \right)^2 \, M^2 \, v_L  + 
 \left( {G_M^p}\, Q \, \right)^2 \, v_T 
  } \ P_2 .
\label{Aperpp2}
\end{eqnarray}

The crucial observation is now that $P_1$ and $P_2$ are related to the 
spin-dependent momentum distributions 
${\cal Y} ( m_3, m_d, m_p ; {\vec q} \, ) $
in the following manner
\begin{eqnarray}
P_1 = 
\frac{ {\cal Y}_1 -  {\cal Y}_2 } 
     { {\cal Y}_1 +  {\cal Y}_2 } ,
\label{p1.1}
\end{eqnarray}
where
\begin{eqnarray}
 {\cal Y}_1 \equiv
{\cal Y} ( m_3=\frac12, m_d=1, m_p=-\frac12 ; {\vec q} \parallel {\hat w} ) 
\nonumber \\
= \frac{1}{12 \, \pi} \, \left(  2 H_0^2 + 2 \sqrt{2} H_0 H_2 + H_2^2  \right)
= \frac{1}{12 \, \pi} \, \left( \sqrt{2} H_0 + H_2  \right)^2 
\label{y1}
\end{eqnarray}
and
\begin{eqnarray}
 {\cal Y}_2 \equiv
{\cal Y} ( m_3=\frac12, m_d=0, m_p= \frac12 ; {\vec q} \parallel {\hat w} )
\nonumber \\
 = \frac{1}{12 \, \pi} \, \left(  H_0^2 - 2 \sqrt{2} H_0 H_2 + 2 H_2^2  \right)
 =  \frac{1}{12 \, \pi} \, \left( H_0 - \sqrt{2} H_2  \right)^2 ,
\label{y2}
\end{eqnarray}
where ${\hat w}$ denotes the spin quantization axis.
Similarly $P_2$ can be written as
\begin{eqnarray}
P_2 = 
\frac{ {\cal Y}_3 +  {\cal Y}_4 - {\cal Y}_5 } 
     { {\cal Y}_3 +  {\cal Y}_4 + {\cal Y}_5 } ,
\label{p1.2}
\end{eqnarray}
where
\begin{eqnarray}
{\cal Y}_3 \equiv {\cal Y} ( m_3=\frac12, m_d=-1, m_p=-\frac12 ; {\vec q} \perp {\hat w} ) 
\ = \ \frac{3}{16 \, \pi} \, H_2^2 , 
\label{y3}
\end{eqnarray}
\begin{eqnarray}
{\cal Y}_4 \equiv {\cal Y} ( m_3=\frac12, m_d= 1, m_p=-\frac12 ; {\vec q} \perp {\hat w} ) 
\nonumber \\
= \frac{1}{48 \, \pi} \, \left(  8 H_0^2 - 4 \sqrt{2} H_0 H_2 + H_2^2  \right) ,
\label{y4}
\end{eqnarray}
and
\begin{eqnarray}
{\cal Y}_5 \equiv {\cal Y} ( m_3=\frac12, m_d= 0, m_p= \frac12 ; {\vec q} \perp {\hat w} ) 
\nonumber \\
= \frac{1}{24 \, \pi} \, \left(  2 H_0^2 + 2 \sqrt{2} H_0 H_2 + H_2^2  \right) .
\label{y5}
\end{eqnarray}

The values of spin projections appearing in Eqs.~(\ref{p1.1}) and (\ref{p1.2})
suggest that $P_1$ and $P_2$ are just the (negative) proton polarizations 
for two different proton momenta ${\vec q}$ inside polarized $^3$He.
To see that this is true we formally define the proton polarization
$P (\vec q \, )$
\begin{eqnarray}
P (\vec q \, ) \equiv
\frac{
\sum\limits_{m_p, m_d} \left| \left\langle \Psi m_3= \frac12 \right|
| \phi_d m_d  \rangle 
| \, {\vec q} \, \frac12 \, m_p \,  \rangle \right|^2 \, m_p 
}
{\frac12 \,
\sum\limits_{m_p, m_d} \left| \left\langle \Psi m_3= \frac12 \right|
| \phi_d m_d  \rangle 
| \, {\vec q} \, \frac12 \, m_p \,  \rangle \right|^2 \,
} .
\label{Pnew}
\end{eqnarray}
Then it is easy to verify that
\begin{eqnarray}
P (\vec q \, \parallel {\hat z} \, ) = -P_1
\label{Pnew1}
\end{eqnarray}
and
\begin{eqnarray}
P (\vec q \, \perp {\hat z} \, ) = -P_2 .
\label{Pnew2}
\end{eqnarray}
We also define the total (integrated) proton polarization as
\begin{eqnarray}
P_{\rm int} & \equiv & 
\frac{\int d{\vec q} \,
\sum\limits_{m_p, m_d} \left| \left\langle \Psi m_3= \frac12 \right|
| \phi_d m_d  \rangle 
| \, {\vec q} \, \frac12 \, m_p \,  \rangle \right|^2 \, m_p 
}
{\frac12 \, \int d{\vec q} \,
\sum\limits_{m_p, m_d} \left| \left\langle \Psi m_3= \frac12 \right|
| \phi_d m_d  \rangle 
| \, {\vec q} \, \frac12 \, m_p \,  \rangle \right|^2 \,
} \nonumber \\
& = & \frac{- \frac{\pi}{24} \, \int\limits_0^\infty d q \, q^2 \, 
\left( \sqrt{2} H_0 + H_2 \, \right)^2 }
{ \frac{\pi}{4} \, \int\limits_0^\infty d q \, q^2 \, 
\left( H_0^2 + H_2^2 \, \right) } 
\equiv \frac{-\int\limits_0^\infty d q \, f_1 (q ) } 
{ \int\limits_0^\infty d q \, f_2 ( q ) }
.
\label{Pnewtotal}
\end{eqnarray}
It is clear that $P_{\rm int}$ is negative. Its numerical value 
obtained with the nuclear forces used in this paper will be given below. 

Thus we can conclude that $P_1$ and $P_2$, 
which can be extracted from the parallel 
and perpendicular helicity asymmetries 
for the $ \vec{^3{\rm He}} ({\vec e},e' p)d$ process, 
if the PWIA approximation is valid, are directly
the proton polarizations inside the polarized $^3$He nucleus.
In the following we will check this simple dynamical 
assumption and compare the results based on the PWIA approximation 
to the results of our full Faddeev calculations. 
We refer the reader to ~\cite{report} for a detailed description of our 
numerical techniques, which we do not want to repeat here.

Note that $P_1$ and $P_2$ are not independent: they are simply related 
since according to Eqs.~(\ref{p1}) and (\ref{p2})
\begin{eqnarray}
2 P_2 = 1 - P_1 .
\label{p1-2}
\end{eqnarray}
If Eqs.~(\ref{Aparap1}) and (\ref{Aperpp2}) are used to obtain the $P_1$ and $P_2$
values from an experiment, then Eq.~(\ref{p1-2}) gives some measure of the validity 
of the PWIA assumption, since the relation (\ref{p1-2}) will in general not hold
for the extracted $P_1$ and $P_2$.

When the argument of $H_0$ and $H_2$ is small ($q \lesssim 50 $ MeV/c), then 
$H_2$ is much smaller than $H_0$. Thus one can expect, quite independent of the details
of the electron kinematics, that
\begin{eqnarray}
P_1 \approx  P_2 \approx \frac13 .
\label{p1-2.2}
\end{eqnarray}

\section{Results for the 
\bm{$ \vec{^3{\rm He}} ({\vec e},e' p)d$} process}
\label{sec:3}

We studied the spin dependent momentum distributions in \cite{spindep} 
and had to conclude that (at least in the nonrelativistic regime) 
one can access these quantities only for rather small pd relative momenta.
The results of \cite{spindep} applied to the 
$ \vec{^3{\rm He}} (e,e' {\vec p})d$ 
and
$ \vec{^3{\rm He}} (e,e' {\vec d})p$ 
processes
but are also valid for 
the $ \vec{^3{\rm He}} ({\vec e},e' p)d$ reaction, 
since the same current matrix elements enter in both calculations.
The important difference is, however, that a measurement 
of the latter reaction, which requires only a polarized electron beam and
a polarized $^3$He target, can be easier realized. 
In fact, this paper is motivated by a very recent experiment ~\cite{Achenbach05},
where for the first time the electron-target asymmetries $A_\parallel$ 
and $A_\perp$ were measured for both the two- and three-body breakup of $^3$He. 
Here we restrict ourselves to one electron kinematics 
from \cite{Achenbach05}
and show its parameters in Table~\ref{tab1}.
\begin{table}[t]
\caption{\label{tab1} Electron kinematics from \cite{Achenbach05}.
E: beam energy, 
$\theta_{e}$: electron scattering angle,
$\omega$: energy transfer,
$Q$: magnitude of the three-momentum transfer $\vec Q$,
$\theta_{Q}$: angle of the three-momentum transfer 
              with respect to the electron beam,
$q^2$: four-momentum transfer squared,
$p_d$: magnitude of the deuteron momentum for proton ejected parallel to $\vec Q$
}
\begin{tabular}{ccccccc}
E & $\theta_{e}$ & $\omega$ & $Q$ & $\theta_{Q}$ & $q^2$ & $p_d$ \\[-8pt]
MeV & deg & MeV  & MeV/c & deg & (GeV/c)$^2$ & MeV/c \\[2pt]
\hline
\ \ \ 735\ \ \  &\ \ \  50\ \ \  &\ \ \  179\ \ \   &\ \ \  569\ \ \  &\ \ \  48.5\ \ \  &\ \ \  0.29\ \ \  &\ \ \  5 \\[2pt]
\end{tabular}
\end{table}

The dynamical input for our calculations
is the nucleon-nucleon force AV18~\cite{AV18} alone or together
with the 3N force UrbanaIX~\cite{UrbanaIX}.
We include in addition to the single nucleon current
the $\pi$- and $\rho$-like two-body currents 
linked to the AV18 force,
following \cite{Riska85} 

\begin{figure}[htb]
\begin{center}
\epsfig{file=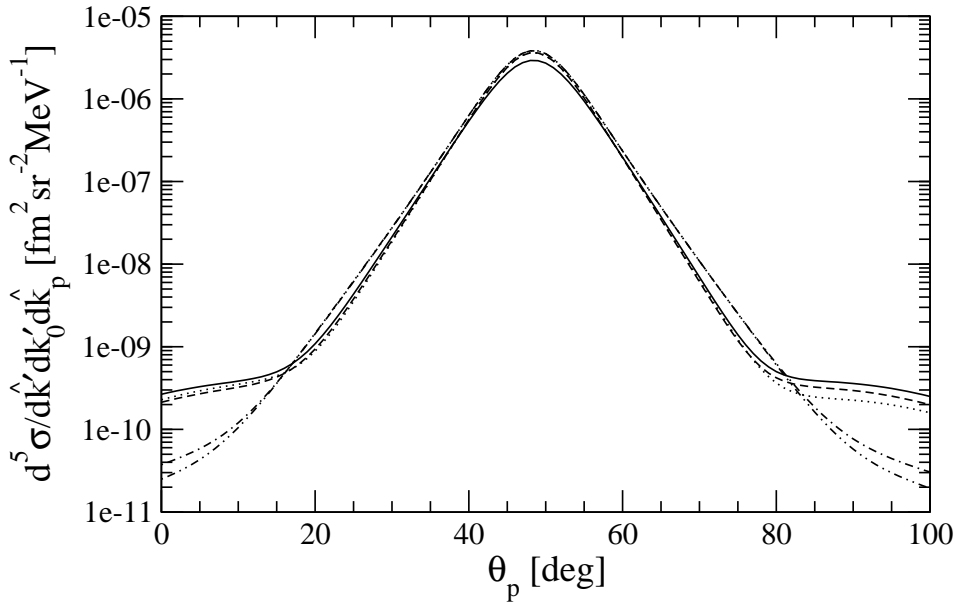,bb=5 350 550 750,clip=true,width=13cm}
\caption{\label{sigma}
Proton angular distribution for the configuration from Table~\ref{tab1}. 
The proton scattering angle $\theta_p$ is defined with respect to the electron beam
so the maximum corresponds to the virtual photon direction $\vec Q$.
The double-dot-dashed curve represents
the prediction based on PWIA.
The dot-dashed curve is obtained under the assumption of PWIAS (which
practically overlaps with PWIA),
the dotted curve takes the full FSI into account but
neglects MEC and 3NF effects.
The $\pi$- and $\rho$-like two-body densities are accounted for additionally
in the dashed curve (which overlaps with FSI),
and finally, the full dynamics including MEC and the 3N force is given
by the solid curve.
      }
\end{center}
\end{figure}

Two-body electron induced breakup of $^3$He is a very rich process.
For example, the description of the deuteron-knockout is not possible 
within the simplest PWIA approximation and complicated rescattering effects 
as well as the details of the nuclear current operator 
play there an important role.
A much simpler dynamical picture is expected 
in the vicinity of the proton knock-out peak. 
We focus on this angular region
and show in Fig.~\ref{sigma} the proton angular distribution
for the selected electron configuration. 
The FSI effects for strictly parallel kinematics amount to 5-7 \%. 
Note that the PWIA results shown in Fig.~\ref{sigma} are obtained without inclusion 
of a 3N force but the full results including the 3N force required both the initial 
and the final state to be calculated with this dynamical component.
The 3N force effects come mainly from the initial bound state
and altogether reach almost 20 \% at $\theta_p = \theta_Q$. 
Note that in this case MEC do not play a big role.

\begin{figure}[htb]
\begin{center}
\epsfig{file=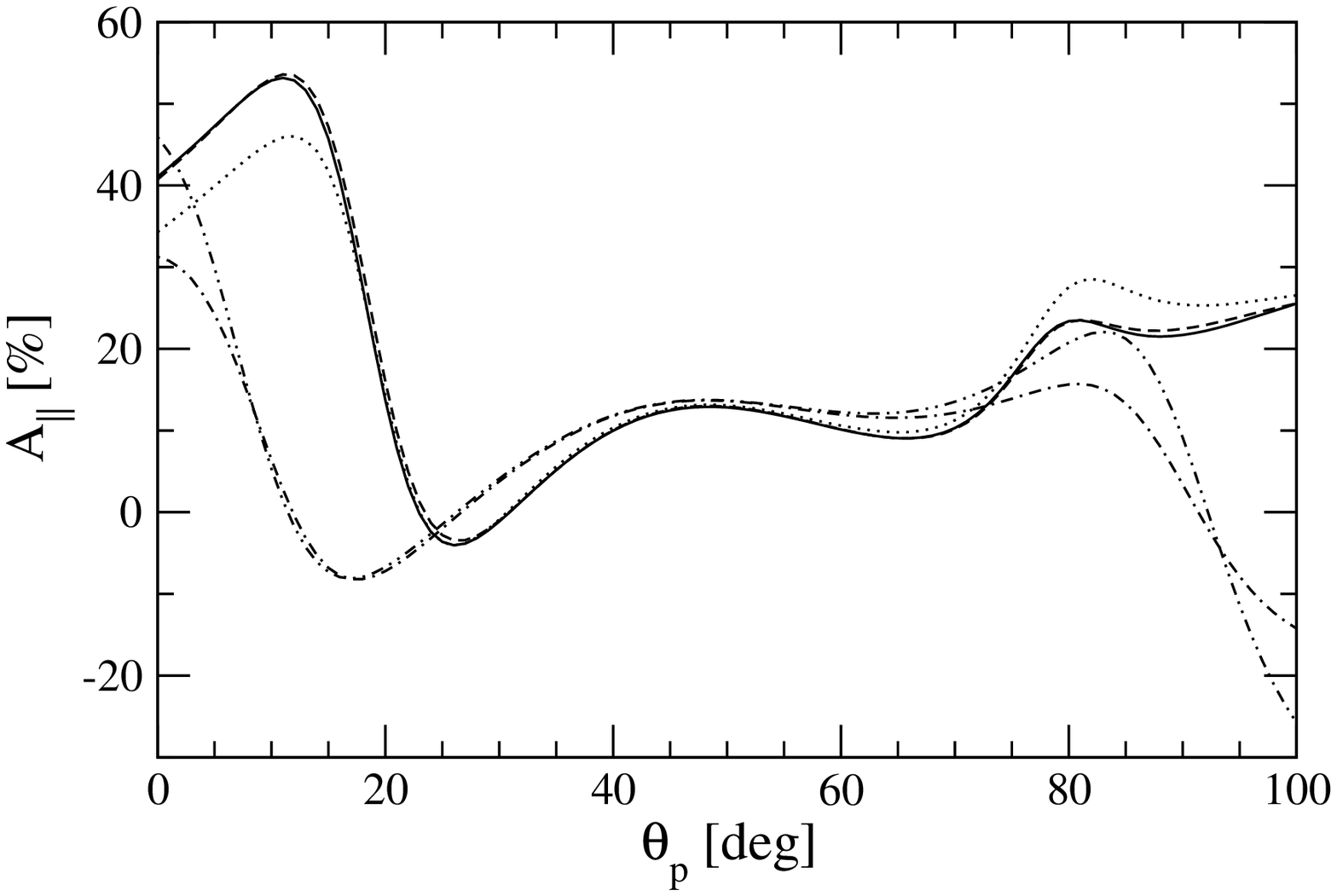,bb=5 350 550 750,clip=true,width=13cm}
\caption{\label{apara.fig}
The parallel helicity asymmetry $A_\parallel$ for the configuration from Table~\ref{tab1}.
Curves as in Fig.~\ref{sigma}.
      }
\end{center}
\end{figure}

\begin{figure}[htb]
\begin{center}
\epsfig{file=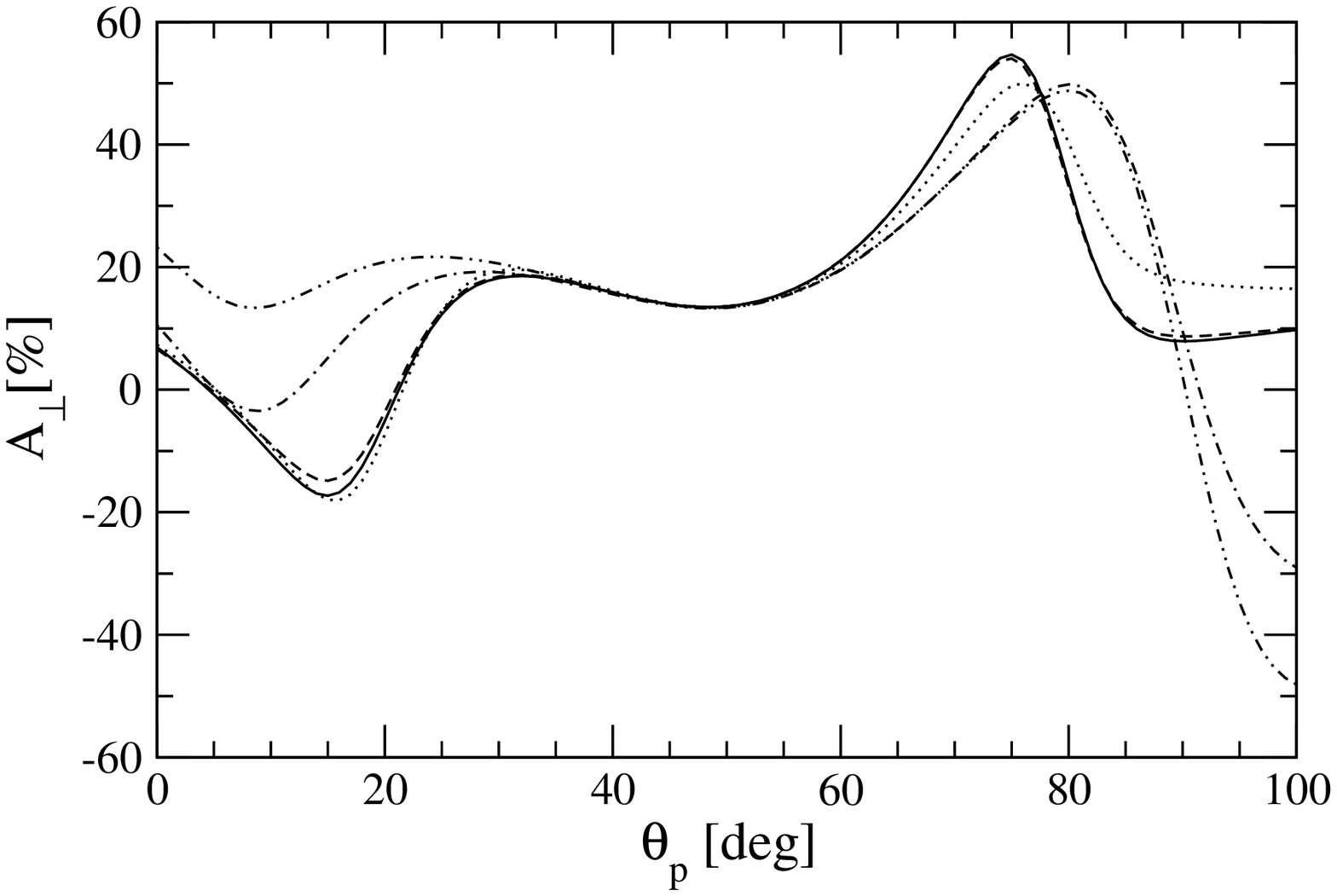,bb=5 350 550 750,clip=true,width=13cm}
\caption{\label{aperp.fig}
The perpendicular helicity asymmetry $A_\perp$ for the configuration from Table~\ref{tab1}.
Curves as in Fig.~\ref{sigma}.
      }
\end{center}
\end{figure}

Let us now turn to the helicity asymmetries shown in Figs.~\ref{apara.fig} and \ref{aperp.fig}.
For $A_\parallel$ the 3N force effects are much smaller 
(below 1 \% for strictly parallel kinematics). FSI are still visible and
slightly reduce the value of $A_\parallel$ in relation to the PWIA result
for parallel kinematics (by nearly 6 \%).

The least sensitivity to the different dynamical ingredients is observed
for $A_\perp$. In Fig.~\ref{aperp.fig} we see that in a certain angular interval 
around $\theta_Q$ all curves overlap. 
That means that in this case one has direct access to important properties
of $^3$He. 

\begin{figure}[htb]
\begin{center}
\epsfig{file=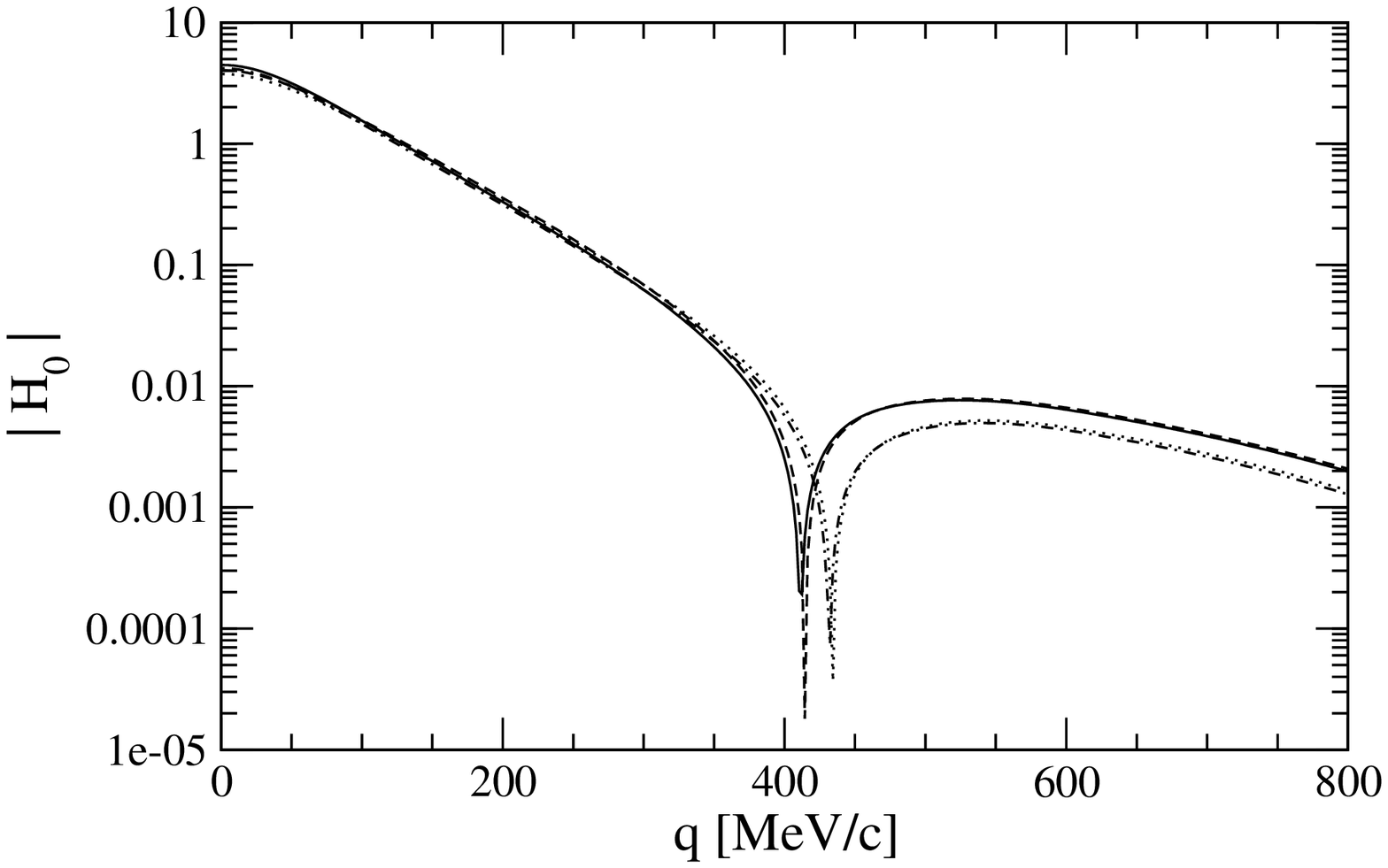,bb=5 350 550 750,clip=true,width=13cm}
\caption{\label{h0.v4.fig}
The quantity $H_0$ for different $^3$He states.
The solid curve corresponds to the full $^3$He state, 
the dashed line shows the results for the case where the S$^\prime$ state
is projected out from the full $^3$He state,
the dotted line represents calculations where only the principal $S$-state
of $^3$He is taken into account and finally 
the dash-dotted line demonstrates the effect of removing the $D$-state
from the full $^3$He wave functions. 
Note that the solid and dashed lines almost completely overlap.
Similarly, the dashed and dash-dotted lines are very close to each other
and are slightly shifted in the zero crossing area. The lack of the $D$-waves 
lowers the magnitudes of $H_0$ at the higher $q$ values.
The underlying full $^3$He wave function was calculated 
including the 3N force.
      }
\end{center}
\end{figure}

\begin{figure}[htb]
\begin{center}
\epsfig{file=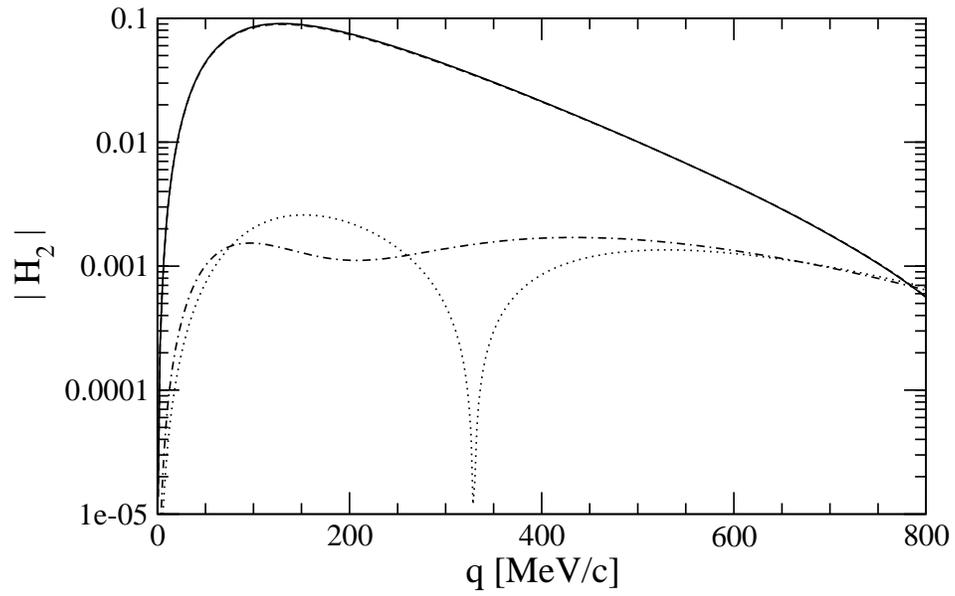,bb=5 350 550 750,clip=true,width=13cm}
\caption{\label{h2.v4.fig}
Curves as in Fig.~\ref{h0.v4.fig} for the quantity $H_2$,
which is clearly dominated by the $D$-state contributions.
      }
\end{center}
\end{figure}

\begin{figure}[htb]
\begin{center}
\epsfig{file=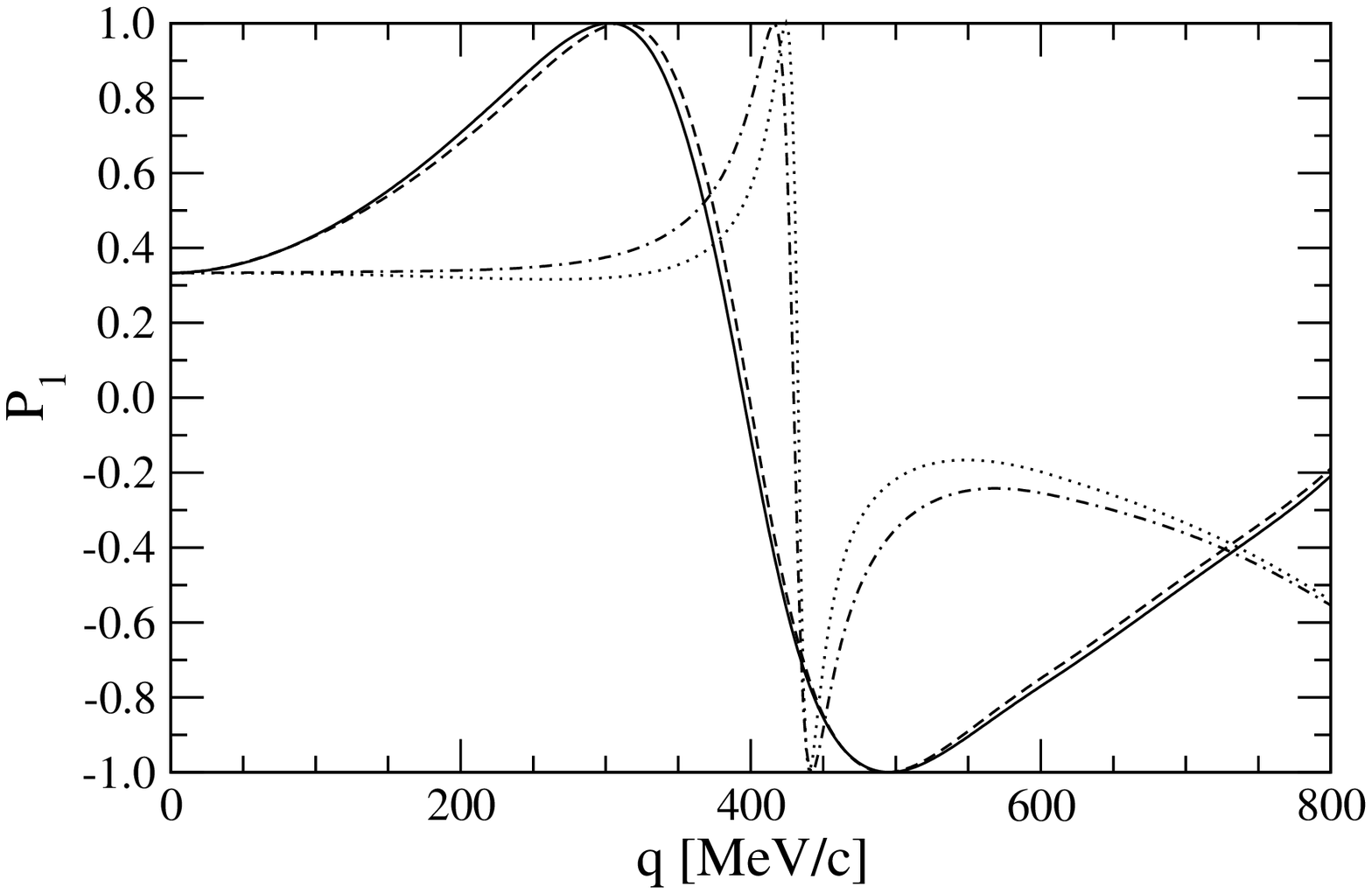,bb=5 350 550 750,clip=true,width=13cm}
\caption{\label{Y1-2.v4.fig}
Curves as in Fig.~\ref{h0.v4.fig} for the quantity $P_1$.
The lack of $D$-wave contributions is clearly visible 
except for very small $q$-values.
      }
\end{center}
\end{figure}

Let us now address the question how (in the given dynamical framework) 
different $^3$He wave function components contribute 
to $H_0$, $H_2$, $P_1$ and $P_2$.
We compare in Figs.~\ref{h0.v4.fig}--\ref{Y1-2.v4.fig} results, 
for the full $^3$He wave function to results obtained with truncated 
wave functions. Besides the full results, we show curves including 
the dominant principal $S$-state, dropping the $D$-
or the $S^\prime$-state contribution. The results, where only 
the principal $S$-state is included, and the ones 
with the $D$-state dropped agree rather well but differ visibly 
from the full prediction.
The neglection of the $S^\prime$-state is hardly noticeable.
The same is true for the $P$-state (not shown).

\begin{figure}[htb]
\begin{center}
\epsfig{file=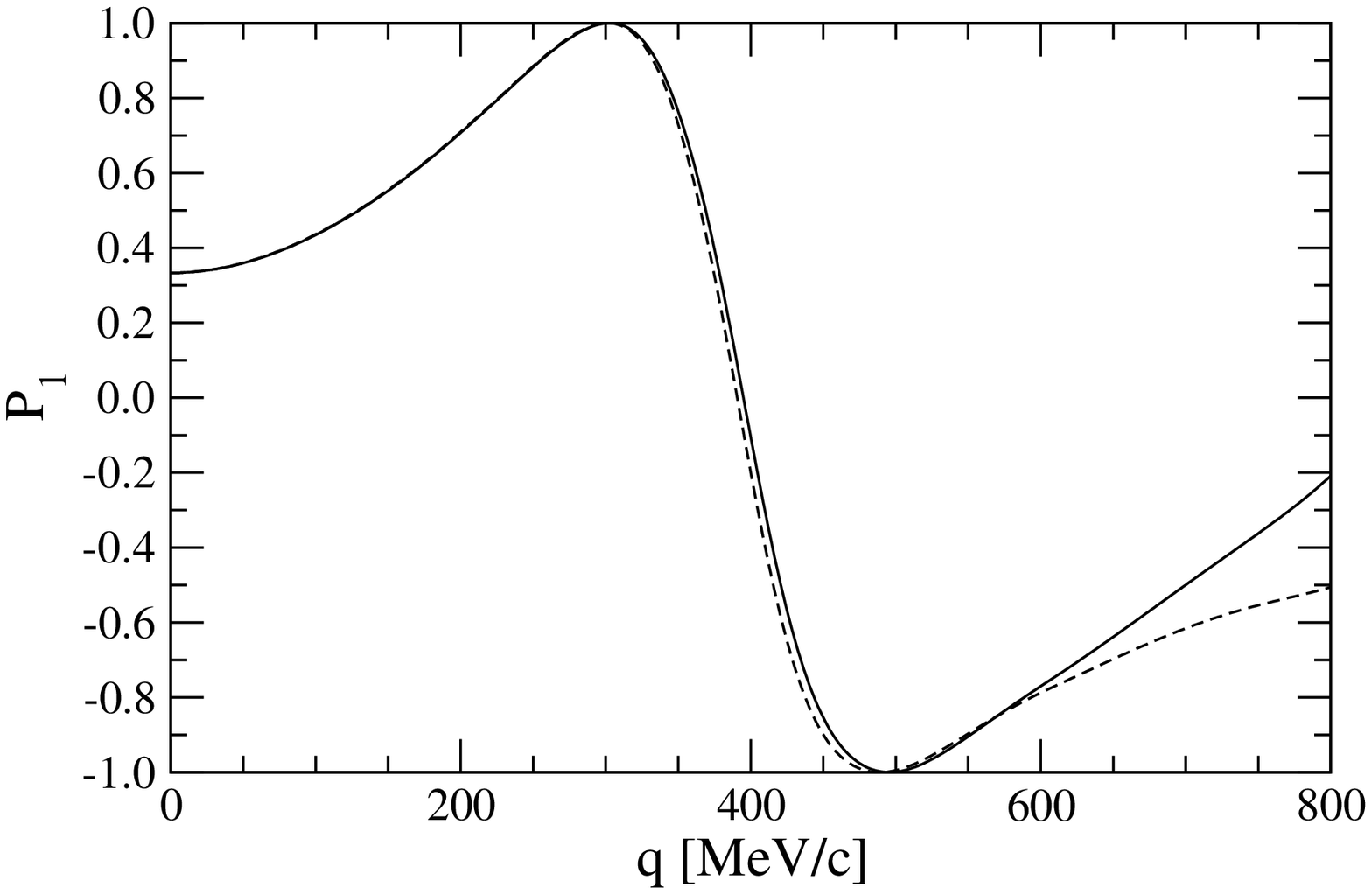,bb=5 350 550 750,clip=true,width=13cm}
\caption{\label{Y1-2.cmp.fig}
The quantity $P_1$ calculated with the inclusion of the 3N force (solid line)
and without the 3N force (dashed line).
      }
\end{center}
\end{figure}

Further we show in Fig.~\ref{Y1-2.cmp.fig}
that the 3N force effects for the quantity $P_1$ 
are rather small. The same holds for $P_2$ (not shown).

\begin{figure}[htb]
\begin{center}
\epsfig{file=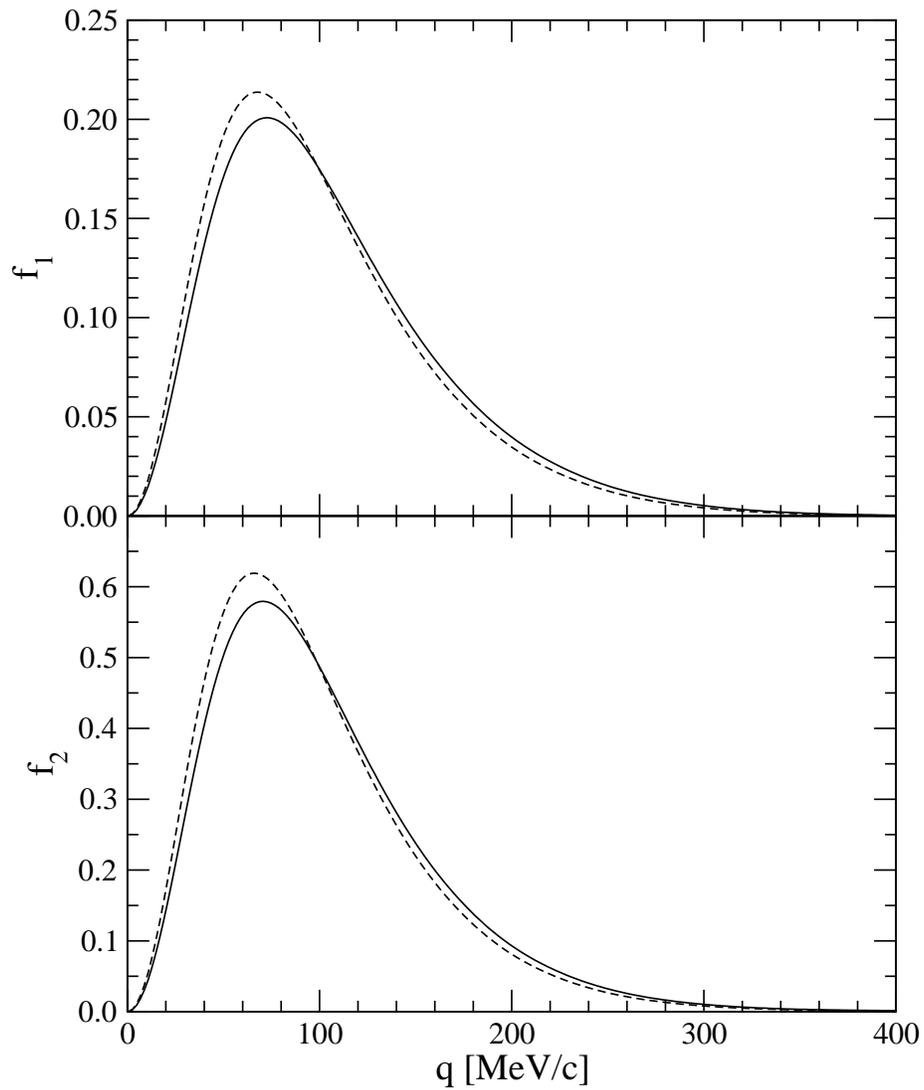,bb=5 50 550 750,clip=true,width=13cm}
\caption{\label{f1-f2.fig}
The integrands $f_1 (q )$ (left) and $f_2 (q ) $ (right) 
appearing in Eq.~(\ref{Pnewtotal}).
Curves as in Fig.~\ref{Y1-2.cmp.fig}.
      }
\end{center}
\end{figure}

We end this section with Fig.~\ref{f1-f2.fig},
which shows the integrands $f_1 (q ) $ and $f_2 (q ) $ 
appearing in the second line of Eq.~(\ref{Pnewtotal}).
We see that relatively small $q$ values ($ q  \lesssim 350 $ (MeV/c)
contribute to $P_{\rm int}$. 
The $P_{\rm int}$ value calculated with (without) the inclusion of the 3N force 
is -0.364 (-0.362).
For completeness we give also the values of the two integrals appearing in Eq.~(\ref{Pnewtotal}):
$ \int\limits_0^\infty d q \, f_1 (q ) $ = 0.127 (0.128),
$ \int\limits_0^\infty d q \, f_2 (q ) $ = 0.348 (0.354)
when calculated with (without) the 3N force. The latter integral gives up to the factor
$\pi/4$ the probability to find a proton-deuteron cluster inside $^3$He. 

\clearpage

\section{Results for the 
\bm{$ \vec{^3{\rm He}} ({\vec e},e' p)pn$} 
and
\bm{$ \vec{^3{\rm He}} ({\vec e},e' n)pp$} 
processes}
\label{sec:4}

In this section the results for the three-body breakup will be discussed.
A general discussion would require that all the elements of our dynamical
framework are involved, i.e. that the initial $^3$He and final scattering 
states are calculated consistently
and many-body currents are taken into account.
We refer the reader to \cite{report} for a discussion 
of the numerical techniques necessary to perform calculations
for such an approach. It, however, precludes any analytical insight.
Thus, as for the $ \vec{^3{\rm He}} ({\vec e},e' p)d$
process, we start with the PWIA approximation. Additionally, we restrict 
the full $^3$He state to its main, principal $S$-state component.
In this case the six nonrelativistic response functions $W_i$ 
for the exclusive 
$ \vec{^3{\rm He}} ({\vec e},e' p)pn$ reaction
take especially simple forms
\begin{eqnarray}
W_L =
\frac{4\,{\,{G_E^p}}^2\,H^2}{6}
\label{wl.ppn}
\end{eqnarray}
\begin{eqnarray}
W_T =
  \frac{2\,H^2\,\left( 
       {\,{G_M^p}}^2\,{\,{Q}}^2 +
       {\,{G_E^p}}^2\,{\,{q_f}}^2 - 
       {\,{G_E^p}}^2\,{\,{q_f}}^2\,\cos (2\,\,{\theta_1})
       \right) }{6\,{\,{M}}^2}
\label{wt.ppn}
\end{eqnarray}
\begin{eqnarray}
W_{TT} =
  \frac{-4\,{\,{G_E^p}}^2\,H^2\,{\,{q_f}}^2\,
     \cos (2\,\,{\phi_1})\,{\sin^2 (\,{\theta_1})}}{6\, {\,{M}}^2}
\label{wtt.ppn}
\end{eqnarray}
\begin{eqnarray}
W_{TL} =
  \frac{8\,{\sqrt{2}}\,{\,{G_E^p}}^2\,H^2\,\,{q_f}\,
     \cos (\,{\phi_1})\,\sin (\,{\theta_1})}{6\,\,{M}}
\label{wtl.ppn}
\end{eqnarray}
\begin{eqnarray}
W_{T^\prime} =  0
\label{wtprime.ppn}
\end{eqnarray}
\begin{eqnarray}
W_{{TL}^\prime} = 0
\label{wtlprime.ppn}
\end{eqnarray}
As before, $\theta_1$ and $\phi_1$ are the polar and azimuthal angles 
corresponding to the $\vec{q}_f \equiv
\frac23 \left[ \vec{p}_1 - \frac12 \left( \vec{p}_2 + \vec{p}_3 \right) \right] 
=  {\vec p}_p - \frac13 {\vec Q} $ direction.
The quantity $H$ is defined as
\begin{eqnarray}
H \, = \,  
\Psi^{\rm PSS} \left( \vec{p}_f , \vec{q}_f - \frac23 \vec{Q} \, \right) ,
\label{HPss}
\end{eqnarray}
where 
$\vec{p}_f$ is the Jacobi momentum describing 
the relative motion within the $23$ (proton-neutron) pair:
\begin{eqnarray}
    \vec{p}_f = \frac12 \left( \vec{p}_2 - \vec{p}_3  \right) .
\label{pf}
\end{eqnarray}
The individual final nucleon momenta are denoted 
by $\vec{p}_1$, $\vec{p}_2$ and $\vec{p}_3$
and the proton, to which the virtual photon is coupled, is the nucleon 1.
The wave function $ \Psi^{\rm PSS} \left( \vec{p} , \vec{q} \, \right)$
is the momentum part of the principal $S$-state:
\begin{eqnarray}
\mid \Psi^{\rm PSS} \rangle \equiv 
\int d^3 p \, \int d^3 q \, 
\mid \vec{p} \, \rangle \, \mid \vec{q} \, \rangle \,
\Psi^{\rm PSS} \left( \vec{p} , \vec{q} \, \right) \, \mid \zeta_a \, \rangle ,
\label{Pss}
\end{eqnarray}
where 
$  \mid \zeta_a \, \rangle $ 
is the completely anti-symmetric 3N spin-isospin state.

The vanishing  of the $ W_{{T}^\prime} $ and $W_{{TL}^\prime} $ 
response functions reflects the well known fact that for the principal 
$S$-state the proton in $^3$He is totally unpolarized.

The situation for the case where the photon ejects the neutron 
is quite different and corresponds very closely to electron scattering  
on a free, fully polarized neutron at rest.
The six nonrelativistic response functions $W_i$ for the 
exclusive 
$ \vec{^3{\rm He}} ({\vec e},e' n)pp$ reaction
under the PWIA approximation and assuming only the principal S-state 
in the $^3$He wave function can be written in the laboratory frame as

\begin{eqnarray}
W_L =
\frac{2\,{\,{G_E^n}}^2\,H^2}{6}
\label{wl.npp}
\end{eqnarray}
\begin{eqnarray}
W_T =
  \frac{H^2\,\left( 
       {\,{G_M^n}}^2\,{\,{Q}}^2 +
       {\,{G_E^n}}^2\,{\,{q_f}}^2 -
       {\,{G_E^n}}^2\,{\,{q_f}}^2\,\cos (2\,\,{\theta_1})
       \right) }{6\,{\,{M}}^2}
\label{wt.npp}
\end{eqnarray}
\begin{eqnarray}
W_{TT} =
  \frac{-2\,{\,{G_E^n}}^2\,H^2\,{\,{q_f}}^2\,
     \cos (2\,\,{\phi_1})\,{\sin^2 (\,{\theta_1})}}{6\, {\,{M}}^2}
\label{wtt.npp}
\end{eqnarray}
\begin{eqnarray}
W_{TL} =
  \frac{4\,{\sqrt{2}}\,{\,{G_E^n}}^2\,H^2\,\,{q_f}\,
     \cos (\,{\phi_1})\,\sin (\,{\theta_1})}{6\,\,{M}}
\label{wtl.npp}
\end{eqnarray}
\begin{eqnarray}
W_{T^\prime} =  
  \frac{-{G_M^n}\,H^2\,\,{Q}\,
       \left( \,{G_M^n}\,\,{Q}\,\cos (\,{\theta^\star}) - 
         2\,\,{G_E^n}\,\,{q_f}\,
          \cos (\,{\phi_1} - \,{\phi^\star})\,
          \sin (\,{\theta_1})\,\sin (\,{\theta^\star}) \right) 
       }{6\,{\,{M}}^2}
\label{wtprime.npp}
\end{eqnarray}
\begin{eqnarray}
W_{{TL}^\prime} = 
  \frac{2\,{\sqrt{2}}\,\,{G_E^n}\,\,{G_M^n}\,H^2\,\,{Q}\,
     \cos (\,{\phi^\star})\,\sin (\,{\theta^\star})}{6\,\,{M}}
\label{wtlprime.npp}
\end{eqnarray}
Now the $\theta_1$ and $\phi_1$ angles 
correspond to the $\vec{q}_f = {\vec p}_1 - \frac13 {\vec Q}
\equiv  {\vec p}_n - \frac13 {\vec Q} $ direction
and $\vec{p}_f$ is the Jacobi momentum describing
the relative motion within the $23$ proton-proton pair.

\begin{figure}[htb]
\begin{center}
\epsfig{file=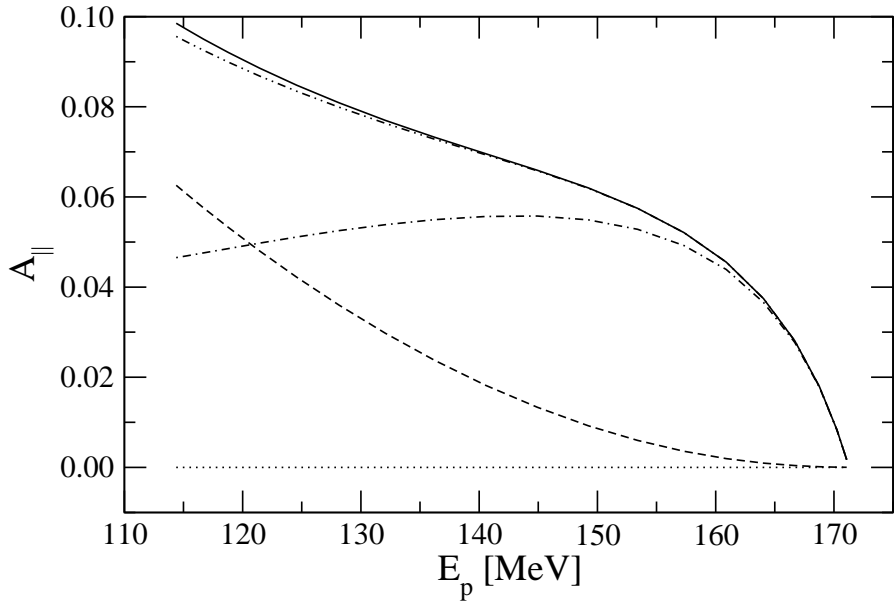,bb=5 350 550 750,clip=true,width=13cm}
\caption{\label{apara.PWIA.ppn}
The parallel asymmetry $A_\parallel$ for the proton ejection in the 
virtual photon direction as a function of the emitted proton energy $E_p$
for the electron configuration from Table~\ref{tab1}
for different $^3$He states.
Curves as in Fig.~\ref{h0.v4.fig} except that the additional double-dot-dashed line
demonstrates the effect of removing the $P$-state
from the full $^3$He wave function.
      }
\end{center}
\end{figure}

\begin{figure}[htb]
\begin{center}
\epsfig{file=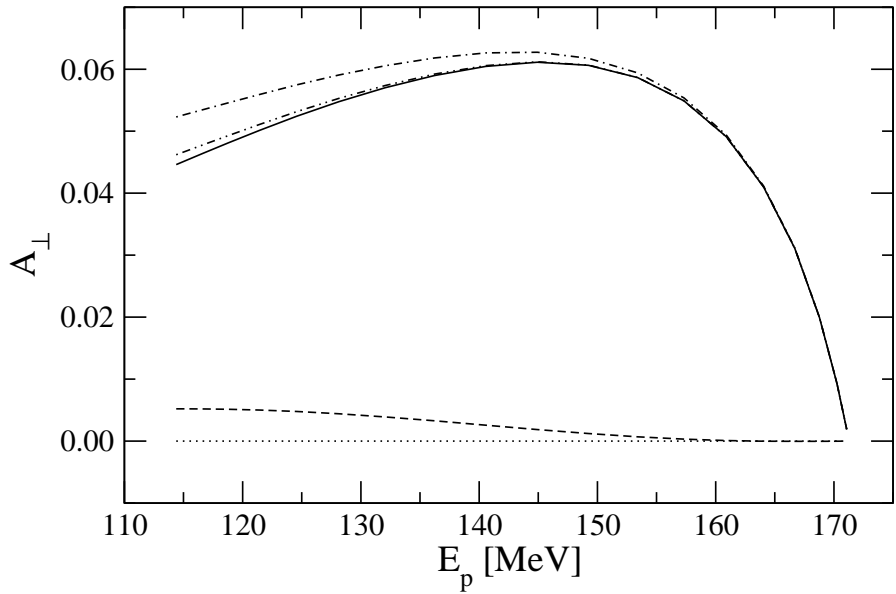,bb=5 350 550 750,clip=true,width=13cm}
\caption{\label{aperp.PWIA.ppn}
The same as in Fig.~\ref{apara.PWIA.ppn} 
for the perpendicular asymmetry $A_\perp$.
      }
\end{center}
\end{figure}

Let us first illustrate the influence of different $^3$He wave function
components on the asymmetries $A_\parallel$ and $A_\perp$
performing calculations that take various components 
of the $^3$He wave function into account.
This is done in Figs.~\ref{apara.PWIA.ppn} and \ref{aperp.PWIA.ppn}
for the $ \vec{^3{\rm He}} ({\vec e},e' p)np$ reaction.
We note that the formulas (\ref{wl.ppn})--(\ref{wtlprime.ppn}) 
and  (\ref{wl.npp})--(\ref{wtlprime.npp}) apply also to the semi-exclusive
reaction. One has to make a simple replacement
\begin{eqnarray}
H^2 \longrightarrow \int d \, {\hat p}_f \, H^2 ,
\label{semi-exclusive}
\end{eqnarray}
i.e., to integrate over the unobserved direction of the 
relative momentum within the $23$ pair.
We see that both asymmetries change quite significantly 
in the given $E_p$ range and become very small for the largest 
$E_p$ values. 
For the principal $S$-state alone both asymmetries are zero. Therefore
the smaller $^3$He components 
(except the $P$-wave) are significant in PWIA
and change the asymmetry in the proton case.
Thereby the $S^\prime$-contribution is more important than the 
$D$-wave piece.

\begin{figure}[htb]
\begin{center}
\epsfig{file=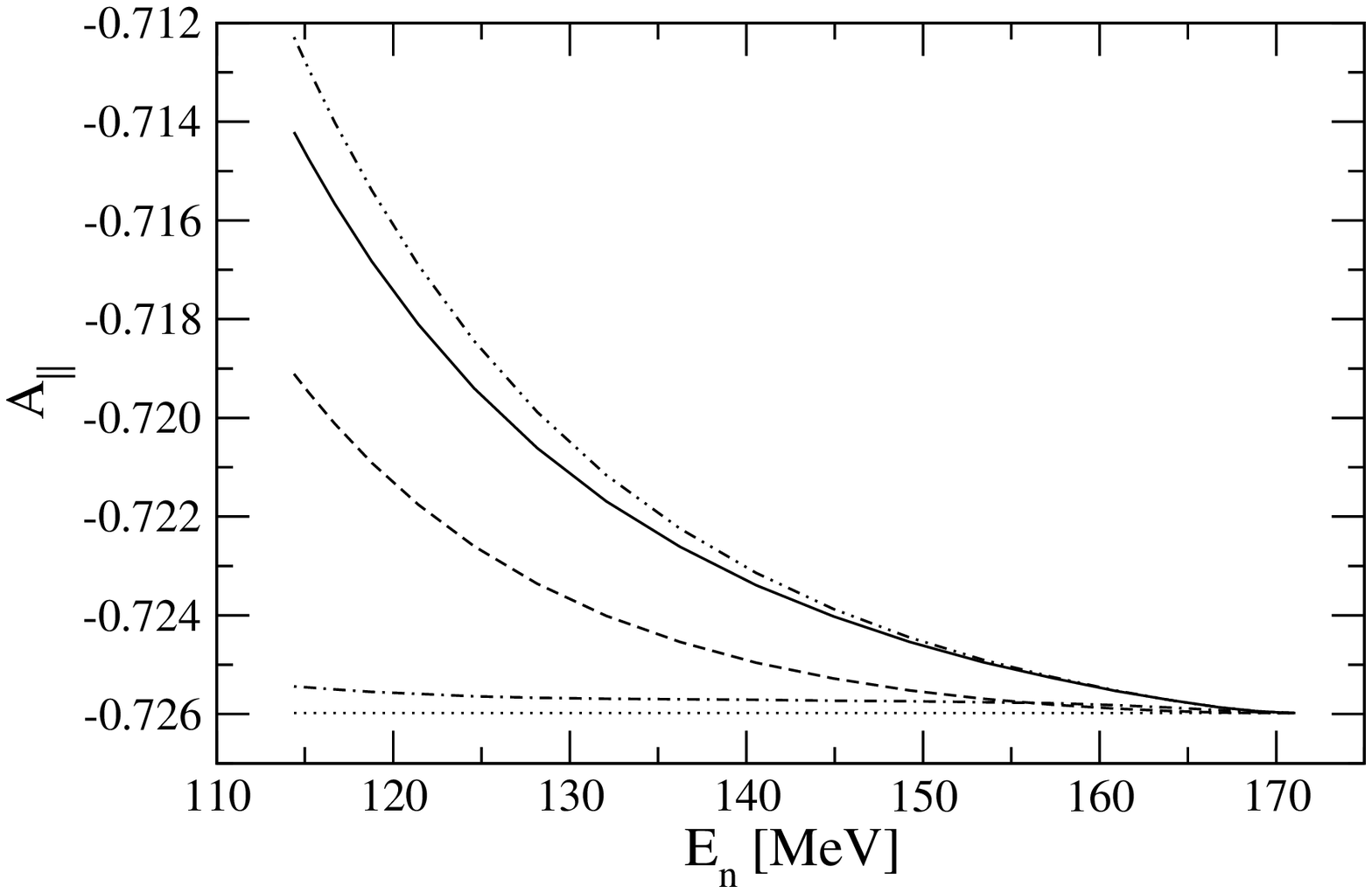,bb=5 350 550 750,clip=true,width=13cm}
\caption{\label{apara.PWIA.npp}
The same as in Fig.~\ref{apara.PWIA.ppn}
for the neutron knock-out.
      }
\end{center}
\end{figure}

\begin{figure}[htb]
\begin{center}
\epsfig{file=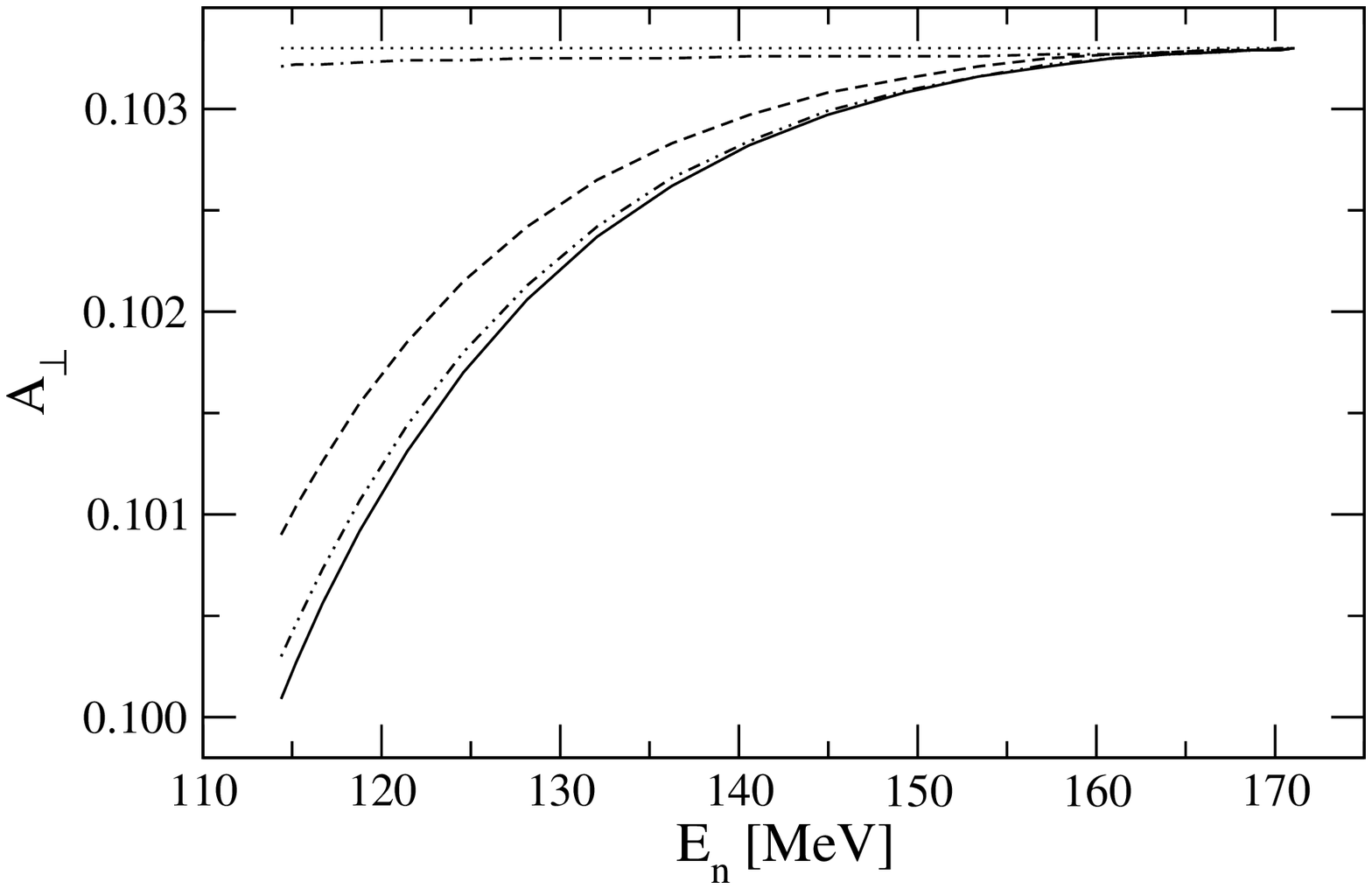,bb=5 350 550 750,clip=true,width=13cm}
\caption{\label{aperp.PWIA.npp}
The same as in Fig.~\ref{aperp.PWIA.ppn}
for the neutron knock-out.
      }
\end{center}
\end{figure}

The situation is quite different for the neutron knock-out 
asymmetries shown in Figs.~\ref{apara.PWIA.npp} and \ref{aperp.PWIA.npp}.
In this case the asymmetries are non-zero even for the principal 
$S$-state wave function.
All results are quite stable in the shown $E_n$ range. 
The change due to different $^3$He states for $A_\parallel$ amounts to 2~\% and 
$A_\perp$ varies by $ \approx $ 3~\%.
The asymmetries reach the specific values which depend only on the neutron 
electromagnetic form factors and trivial kinematic factors $v_i$
appearing in Eq.~(\ref{eq4}).

The PWIA picture is very simple but quite unrealistic. 
That is why FSI has to be taken into account. In order to retain analytical 
insight but make our framework more realistic we will in the following 
additionally allow for the rescattering effects in the subsystem $23$ and  
call the resulting approximation $FSI23$. 
The $^3$He wave function will still be restricted to the principal $S$-state.
Thus we need the following matrix elements $N^\mu$
of the single nucleon current $j_1^\nu ( {\vec Q} )$
(see \cite{report} for more details of our notation)
\begin{eqnarray}
N^\mu =
\langle \, 
{\vec p}_1 \, 
{\vec p}_{23} \, 
{\vec p}_f \, 
s m_s \, \frac12 m_1 ; t m_t \, \frac12 \nu_1 \mid 
\left( 1 + t_{23} G_0 \right) j_1^\nu ( {\vec Q} )
 \mid  \Psi^{\rm PSS} M_S M_T = \frac12 \rangle ,
\label{fsi23}
\end{eqnarray}
where 
\begin{eqnarray}
{\vec p}_{23} \equiv {\vec p}_2 + {\vec p}_3 , 
\label{p23}
\end{eqnarray}
the $23$ pair spin and spin projection are
denoted by $s$ and $m_s$, 
the $23$ pair isospin and isospin projection are 
$t$ and $m_t$ and the spin and isospin magnetic quantum numbers 
of the nucleon~1 are $m_1$ and $\nu_1$.
The total $^3$He spin and isospin projections are $M_S$ and $M_T$,
respectively. Further $t_{23}$ is the NN $t$-matrix acting within the $23$ pair
and $G_0$ is the free 3N propagator.

The six nonrelativistic response functions $W_i$ for the
exclusive
$ \vec{^3{\rm He}} ({\vec e},e' p)pn$
reaction
under the FSI23 approximation and assuming only the principal S-state
in the $^3$He wave function can be written in the laboratory frame as

\begin{eqnarray}
W_L = \frac{ {G_E^p}^2\, H_1 }{6}
\label{wl.ppn.fsi23}
\end{eqnarray}
\begin{eqnarray}
W_T = 
\frac{\left( {\,{G_M^p}}^2\,{\,{Q}}^2 + 
      {\,{G_E^p}}^2\,{\,{q_f}}^2 - 
      {\,{G_E^p}}^2\,{\,{q_f}}^2\,\cos (2\,\,{\theta_1}) \
\right) \, H_1 \,
}{12\,{\,{M}}^2}
\label{wt.ppn.fsi23}
\end{eqnarray}
\begin{eqnarray}
W_{TT} =
\frac{- {\,{G_E^p}}^2\,{\,{q_f}}^2\,
      \cos (2\,\,{\phi_1})\,
\,{\sin^2 (\,{\theta_1})} \, H_1 \, }{6\, {\,{M}}^2}
\label{wtt.ppn.fsi23}
\end{eqnarray}
\begin{eqnarray}
W_{TL} =
\frac{{\sqrt{2}}\,{\,{G_E^p}}^2\,\,{q_f}\,
    \cos (\,{\phi_1})\, 
\,\sin (\,{\theta_1}) \, H_1 \, }{3\,\,{M}}
\label{wtl.ppn.fsi23}
\end{eqnarray}
\begin{eqnarray}
W_{T^\prime} & = &
\frac{\, 
-\left( {G_M^p}\,{Q}\, \right)^2 \, H_2 \,  \cos (\,{\theta^\star})\, }{12\,{\,{M}}^2} 
\nonumber \\
& + & \frac{\, 
2\,{\sqrt{2}}\,\,{G_E^p}\, {G_M^p}\,{Q}\, {q_f}\, H_3 \, \cos (\,{\phi_1})\, \sin (\,{\theta_1} )
   \, \cos (\,{\theta^\star}) \,}
{12\,{\,{M}}^2}
\nonumber \\
& + & \frac{\, 
2\,{\sqrt{2}}\,\,{G_E^p}\, {G_M^p}\,{Q}\, {q_f}\, i \, H_4 \, \sin (\,{\phi_1})\, \sin (\,{\theta_1}  )
\, \cos (\,{\theta^\star}) \,}
{12\,{\,{M}}^2}
\nonumber \\
& - &
\frac{ 
\, {\sqrt{2}}\, \left( {G_M^p}\,\,{Q}\, \right)^2 \,
          \left( \cos (\,{\phi^\star})\, \, H_3 \, + i \, H_4 \, \sin (\,{\phi^\star}) \right) 
  \,\sin (\,{\theta^\star})
}
{12\,{\,{M}}^2} 
\nonumber \\
& + &
\frac{\, 2 \, {G_E^p}\, {G_M^p}\,{Q}\, {q_f}\, H_5 \, \cos (\,{\phi_1}  + \,{\phi^\star})\, 
            \sin (\,{\theta_1}) \,\sin (\,{\theta^\star}) }{12\,{\,{M}}^2} 
\nonumber \\
& + & 
\frac{\, 2 \, {G_E^p}\, {G_M^p}\,{Q}\, {q_f}\, H_6 \, \cos (\,{\phi_1} - \,{\phi^\star})\, 
            \sin (\,{\theta_1}) \,\sin (\,{\theta^\star}) }{12\,{\,{M}}^2} 
\nonumber \\
& - & 
\frac{\, 2 \, {G_E^p}\, {G_M^p}\,{Q}\, {q_f}\, i \, H_7 \, \sin (\,{\phi_1} + \,{\phi^\star})\, 
            \sin (\,{\theta_1}) \,\sin (\,{\theta^\star}) }{12\,{\,{M}}^2} 
\label{wtprime.ppn.fsi23}
\end{eqnarray}
\begin{eqnarray}
W_{TL^\prime} & = &
\frac{\, 2 \, {G_E^p}\,\,{G_M^p}\,\,{Q}\, H_3 \, \cos (\,{\theta^\star})\, }{6\, \,{M}}
\nonumber \\
& + &
\frac{\,  {\sqrt{2}}\, {G_E^p}\,\,{G_M^p}\,\,{Q}\, H_8 \, \cos (\,{\phi^\star})\, \sin (\,{\theta^\star}) \, }{6\, \,{M}}
\nonumber \\
& - & 
\frac{\, 4\,{\sqrt{2}}\, {G_E^p}\,\,{G_M^p}\,\,{Q}\, H_9 \,  \sin (\,{\phi^\star})\,\sin (\,{\theta^\star}) \, }{6\, \,{M}}
\label{wtlprime.ppn.fsi23}
\end{eqnarray}
The auxiliary quantities $H_1$--$H_9$ are
\begin{eqnarray}
H_1 \ \equiv \
{\mid G(1) \mid}^2 + 
      2\,\left( {\mid G(4) \mid}^2 + {\mid G(5) \mid}^2 + 
         {\mid G(6) \mid}^2 + {\mid G(7) \mid}^2 \right)  + 
      {G(8)}^2
\label{H1}
\end{eqnarray}
\begin{eqnarray} 
H_2 \ \equiv \
       {\mid G(1) \mid }^2 - 
              2\,\left( {\mid G(4) \mid }^2 - 
                 {\mid G(5) \mid }^2 + {\mid G(6) \mid }^2 + 
                 {\mid G(7) \mid }^2 \right)  + {G(8)}^2 
\label{H2}
\end{eqnarray}
\begin{eqnarray}
H_3 \ \equiv \
             \left( (G(4))^\star - (G(6))^\star \right) \,G(5) + 
               (G(5))^\star\, \left( G(4) - G(6) \right)  + 
               2\,G(8)\, \Re {G(7)} 
\label{H3}
\end{eqnarray}
\begin{eqnarray}
H_4 \ \equiv \
- \left(  \left( (G(4))^\star + (G(6))^\star \right) \,G(5) \ \right)  
+ (G(5))^\star \, \left( G(4) + G(6) \right)  +  2\, i \,G(8)\, \Im {G(7)} 
\label{H4}
\end{eqnarray}
\begin{eqnarray}
H_5 \ \equiv \
              {(G(7))^\star}^2 
             - 2\, (G(6))^\star\,G(4) 
             - 2\,(G(4))^\star \,G(6) 
             + {G(7)}^2 
\label{H5}
\end{eqnarray}
\begin{eqnarray}
H_6 \ \equiv \
              { \mid G(1) \mid }^2 
              - 2\,{ \mid G(5) \mid }^2 
              - {G(8)}^2 
\label{H6}
\end{eqnarray}
\begin{eqnarray}
H_7 \ \equiv \
               {(G(7))^\star}^2 
               + 2\,(G(6))^\star \,G(4) 
               - 2\,(G(4))^\star\,G(6) 
               - {G(7)}^2 
\label{H7}
\end{eqnarray}
\begin{eqnarray}
H_8 \ \equiv \
         { \mid G(1) \mid }^2 
         - 2\,{\mid G(5) \mid }^2 
         + {(G(7))^\star}^2
          \nonumber \\
         - 2\,(G(6))^\star\,G(4) 
         - 2\,(G(4))^\star \,G(6) 
         + {G(7)}^2 - {G(8)}^2 \
\label{H8}
\end{eqnarray}
\begin{eqnarray}
H_9 \ \equiv \
         \Im {G(6)}\,\Re {G(4)} 
       - \Im {G(4) }\,\Re {G(6)} 
       + \Im {G(7) }\,\Re {G(7)} 
\label{H9}
\end{eqnarray}
The different $G(i)$ functions that appear in the equations 
are the integrals 
\begin{eqnarray}
F ( s ,  m_s , m_{s^\prime} , t , m_t ) \ \equiv \ 
\int d {\vec p}^{\, \prime} \, 
\langle {\vec p}_f \, s m_s \, t m_t \mid 1 + t_{23} G_0 \mid 
{\vec p}^{\, \prime} \, s m_{s^\prime} \, t m_t \rangle  \,
\Psi^{\rm PSS} \left( {\vec p}^{\, \prime} , \vec{q}_f - \frac23 \vec{Q} \, \right)
\label{Fs}
\end{eqnarray}
for different combinations of $s$, $m_s$, $m_{s^\prime}$, $t$ and $m_t$:
\begin{eqnarray}
G (1) = F ( 0 , 0 , 0 , 1 , 0 )\nonumber   \\
G (2) = F ( 0 , 0 , 0 , 1 , 1 )\nonumber   \\
G (3) = F ( 0 , 0 , 0 , 1 , -1 ) \nonumber  \\
G (4) = F ( 1 , -1 , -1 , 0 , 0 ) \nonumber  \\
G (5) = F ( 1 , -1 , 0 , 0 , 0 ) \nonumber  \\
G (6) = F (  1 , -1 , 1 , 0 , 0 ) \nonumber  \\
G (7) = F ( 1 , 0 , -1 , 0 , 0 ) \nonumber  \\
G (8) = F (  1 , 0 , 0 , 0 , 0 ) \nonumber  \\
G (9) = F (  1 , 0 , 1 , 0 , 0 ) \nonumber  \\
G (10) = F (  1 , 1 , -1 , 0 , 0 ) \nonumber  \\
G (11) = F (  1 , 1 , 0 , 0 , 0 ) \nonumber  \\
G (12) = F (  1 , 1 , 1 , 0 , 0 )
\end{eqnarray}
In the case of $^3$He $ G (3) $ is absent.

Due to the assumed $t$-matrix properties (isospin invariance 
and invariance with respect to time reversal)
\begin{eqnarray}
G(3) = G(2) = G(1) \nonumber  \\
G(10) = \left( G(6) \right)^\star \nonumber  \\
G(9) = -\left( G(7) \right)^\star \nonumber  \\
G(12) = \left( G(4) \right)^\star \nonumber  \\
G(11) = -\left( G(5) \right)^\star \nonumber  \\
G(8) = \left( G(8) \right)^\star 
\end{eqnarray}
some of the combinations could be eliminated. 
When the term $  t_{23} G_0 $ in Eq.~(\ref{Fs}) is dropped then 
\begin{eqnarray}
F ( s ,  m_s , m_{s^\prime} , t , m_t ) = \delta_{ m_s , m_{s^\prime} } \, ,
\end{eqnarray}
the quantities $H_2$--$H_9$ vanish
and $H_1$ reduces to $ 4 \left( G(1) \right)^2 $. In this way the PWIA results
of Eqs.~(\ref{wl.ppn})--(\ref{wtlprime.ppn}) are recovered.

For the sake of completeness we give also the corresponding 
(and much simpler) expressions for the six nonrelativistic 
response functions $W_i$ in the
case of the exclusive
$ \vec{^3{\rm He}} ({\vec e},e' n)pp$ reaction
under the same dynamical assumptions:
\begin{eqnarray}
W_L= \frac{{\,{G_E^n}}^2\,{\mid G(2) \mid}^2}{3}
\label{wl.npp.pss.fsi23} 
\end{eqnarray}
\begin{eqnarray}
W_T= \frac{{\mid G(2) \mid}^2\,
    \left( {\,{G_M^n}}^2\,{\,{Q}}^2 + 
      {\,{G_E^n}}^2\,{\,{q_f}}^2 - 
      {\,{G_E^n}}^2\,{\,{q_f}}^2\,\cos (2\,\,{\theta_1}) \
\right) }{6\,{\,{M}}^2}
\label{wt.npp.pss.fsi23} 
\end{eqnarray}
\begin{eqnarray}
W_{TT}= \frac{-  {\mid G(2) \mid}^2\, 
          {\,{G_E^n}}^2\,{\,{q_f}}^2\,
      \cos (2\,\,{\phi_1})\,
      {\sin^2 (\,{\theta_1})} }{3\,{\,{M}}^2}
\label{wtt.npp.pss.fsi23} 
\end{eqnarray}
\begin{eqnarray}
W_{TL}=
\frac{2\,{\sqrt{2}}\,{\,{G_E^n}}^2\,\,{q_f}\,
    {\mid G(2) \mid}^2\,\cos (\,{\phi_1})\,
    \sin (\,{\theta_1})}{3\,\,{M}}
\label{wtl.npp.pss.fsi23} 
\end{eqnarray}
\begin{eqnarray}
W_{T^\prime} =
\frac{- {G_M^n}\,\,{Q}\,{\mid G(2) \mid}^2\,
      \left( \,{G_M^n}\,\,{Q}\,\cos (\,{\theta^\star}) - 
        2\,\,{G_E^n}\,\,{q_f}\,
         \cos ({\phi_1} - \,{\phi^\star})\,
         \sin ({\theta_1})\,\sin ({\theta^\star}) \,
\right) }{6\,{\,{M}}^2}
\label{wtprime.npp.pss.fsi23} 
\end{eqnarray}
\begin{eqnarray}
W_{TL^\prime} =
\frac{{\sqrt{2}}\,\,{G_E^n}\,\,{G_M^n}\,\,{Q}\,
    {\mid G(2) \mid}^2\,\cos (\,{\phi^\star})\,
    \sin (\,{\theta^\star})}{3\,\,{M}}
\label{wtlprime.npp.pss.fsi23} 
\end{eqnarray}
The response functions have the same form as for the PWIA 
approximation displayed in Eqs.~(\ref{wl.npp})--(\ref{wtlprime.npp}).
The simple form of 
Eqs.~(\ref{wl.npp.pss.fsi23})--(\ref{wtlprime.npp.pss.fsi23})
is guaranteed by the fact that for the neutron emission only $t$-matrices 
with the total subsystem isospin $t=1$ contribute.
If one forms now the helicity asymmetries 
$A (\theta^\star \, , \, \phi^\star \, )$,
then exactly the same form is obtained as in the case of PWIA, i.e.,
all information from $^3$He (restricted to the principal $S$-state)
disappears.

The formula (\ref{fsi23}) and the following $t$-matrices are given in 
the three-vector representation. Since we work with partial wave decomposed 
$t$-matrices, it is adequate to ask the question if the interaction is 
dominated by one or very few channel states. Further we would like to see 
if the truncation of the $^3$He wave function to the principal $S$-state
is reasonable, at least for the highest energies of the emitted nucleon.

\begin{figure}[htb]
\begin{center}
\epsfig{file=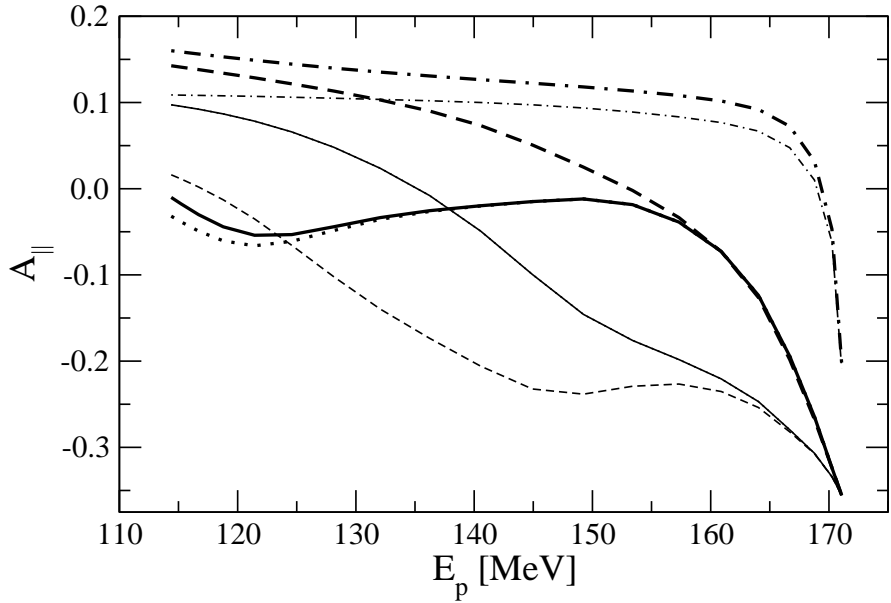,bb=5 350 550 750,clip=true,width=13cm}
\caption{\label{apara.fsi23}
The parallel asymmetry $A_\parallel$ for the proton ejection in the
virtual photon direction as a function of the ejected proton energy $E_p$
for the electron configuration from Table~\ref{tab1} under 
the $FSI23$ approximation.
Dash-dotted lines are obtained for the case, where the $t$-matrix acts only 
in the $^1S_0$ channel, dashed lines correspond to the calculations 
in which only the $^1S_0$ and $^3S_1$ $t$-matrix components are taken into account 
(without coupling to the  $^3D_1$ state), dotted lines show the results 
for $^1S_0$ and $^3S_1$--$^3D_1$ states and finally solid lines 
correspond to inclusion of all nucleon-nucleon $t$-matrix partial waves 
with the total angular momentum $j \le 3$. Thick lines are
obtained with the full $^3$He state and thin lines with $^3$He
truncated to the principal $S$-state.
Note that the thin dotted and solid lines completely overlap.
      }
\end{center}
\end{figure}

\begin{figure}[htb]
\begin{center}
\epsfig{file=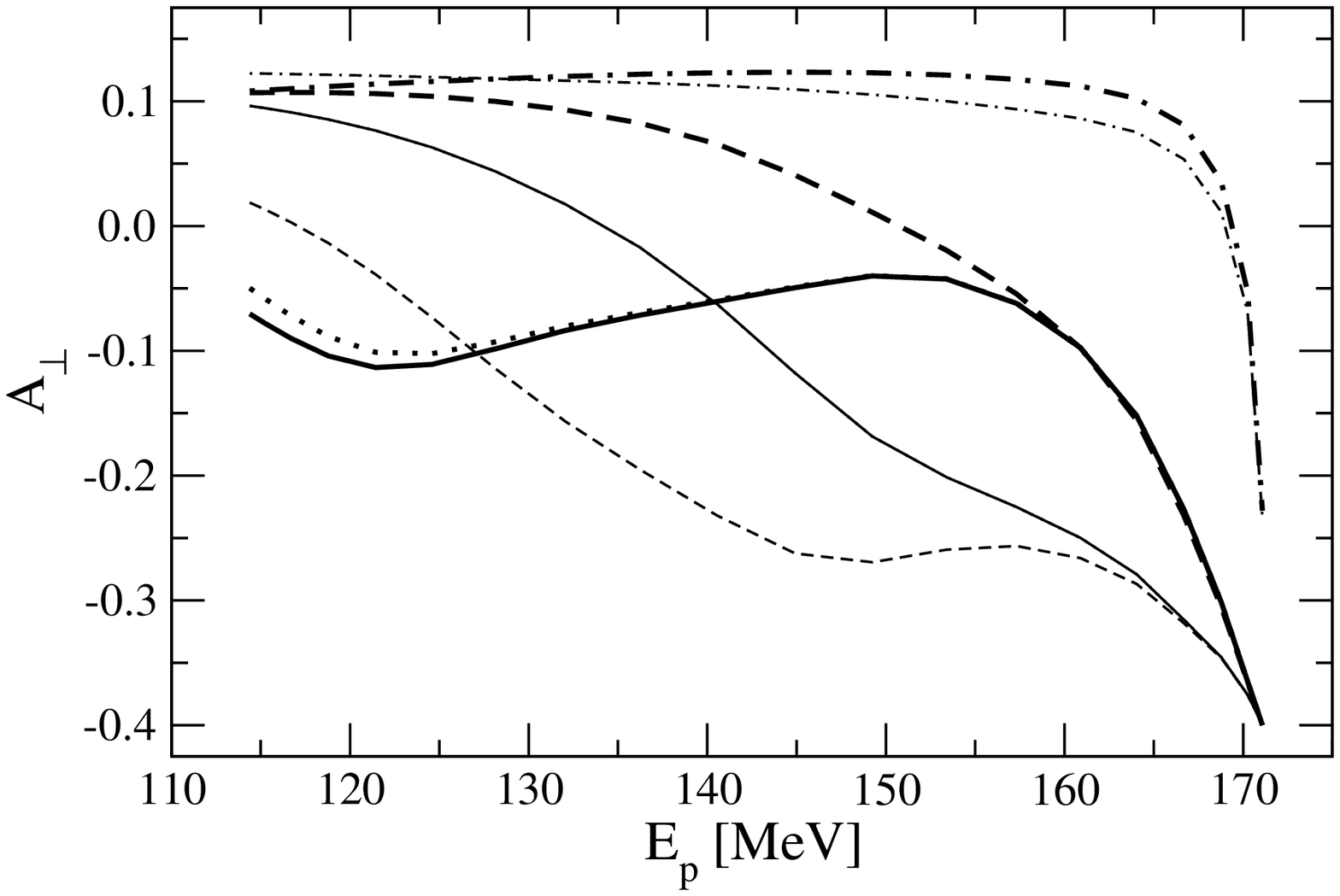,bb=5 350 550 750,clip=true,width=13cm}
\caption{\label{aperp.fsi23}
The same as in Fig.~\ref{apara.fsi23} for the perpendicular asymmetry $A_\perp$.
      }
\end{center}
\end{figure}

Let us start with the more intricate case of the proton emission.
In Figs.~\ref{apara.fsi23} and \ref{aperp.fsi23} we show different curves
obtained with the full $^3$He state (thick lines) and with $^3$He 
truncated to the principal $S$-state (thin lines) for different 
number of $t$-matrix partial waves. 
We note first of all that both cases of the parallel and perpendicular asymmetries 
are quite similar, especially for the range of the asymmetry values.
It is clear that the truncation of the full $^3$He wave function to the principal $S$-state 
is valid only for the highest emission energies. Otherwise the influence of the smaller 
$^3$He  wave function components is very strong. 
Another important observation is that even for these highest energies 
the action of the $t$-matrix cannot be restricted to just one $^1S_0$ channel
and the inclusion at least of the $^3S_1$ partial wave state is inevitable. 
Since then both spins $s=0$ and $s=1$ appear for the $np$ subsystem, 
the photon couples to the proton which is polarized along and opposite
to the spin of polarized  $^3$He. If in the $np$ subsystem only 
the spin $s=0$ were active, the photon would couple to the 100~\% 
polarized proton.

\begin{figure}[htb]
\begin{center}
\epsfig{file=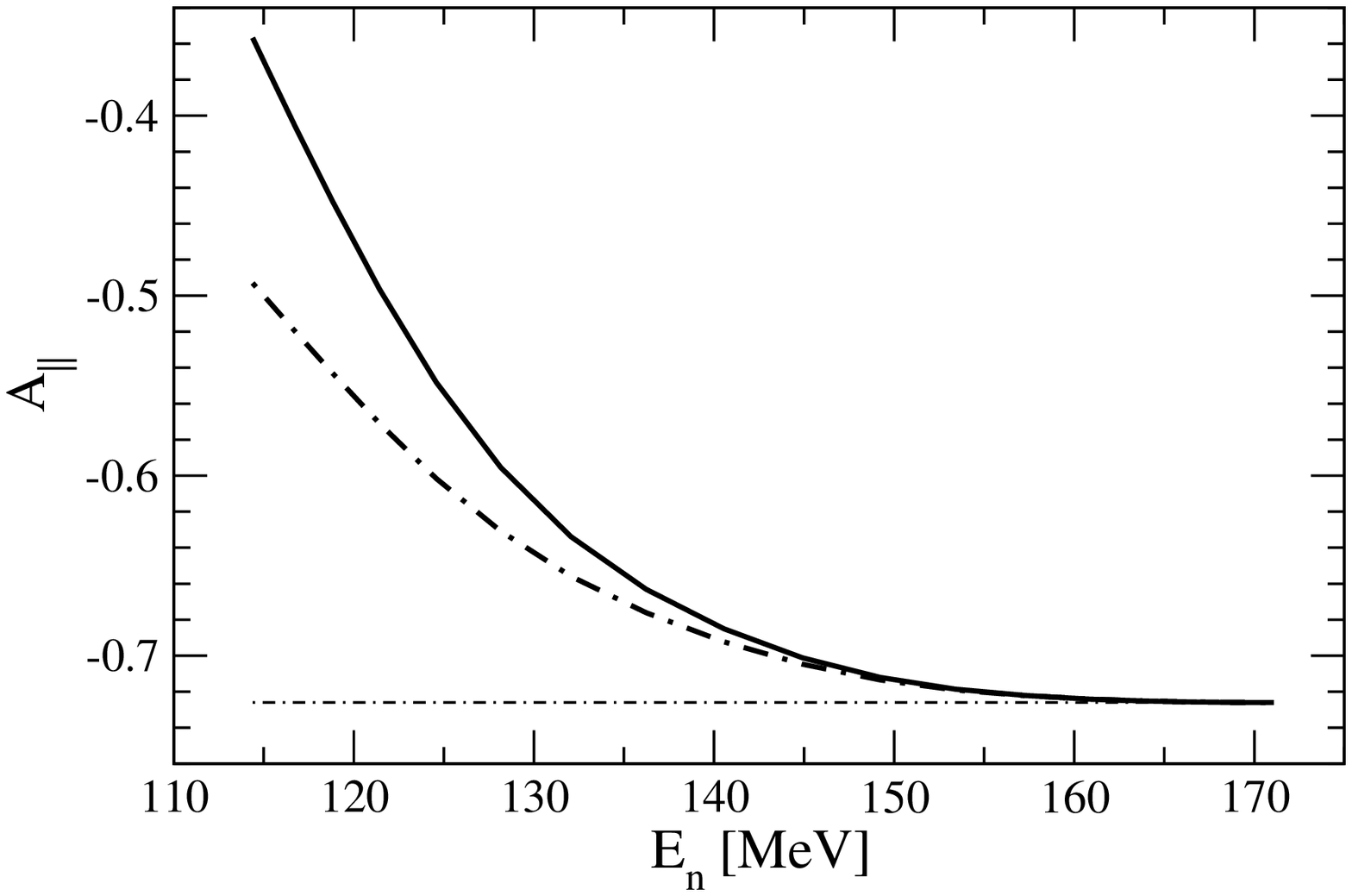,bb=5 350 550 750,clip=true,width=13cm}
\caption{\label{apara.fsi23.npp}
The parallel asymmetry $A_\parallel$ for the neutron emission
in the virtual photon direction as a function of the ejected neutron energy $E_n$
for the electron configuration from Table~\ref{tab1} under
the $FSI23$ approximation.
Since in this case only $t=1$ states contribute to the $t_{23} G_0 $ part
of the $FSI23$ matrix elements, we show only three cases:
the results obtained with the principal $S$-state and the $t$-matrix
restricted to the $^1S_0$ state (thin dash-dotted line),
the results obtained with the full $^3$He wave function and the $t$-matrix
restricted to the $^1S_0$ state (thick dash-dotted line),
and finally the results obtained with the full $^3$He wave function and the $t$-matrix
acting in all partial waves with the total angular momentum $ j \le 3$ 
(thick solid line).
      }
\end{center}
\end{figure}

\begin{figure}[htb]
\begin{center}
\epsfig{file=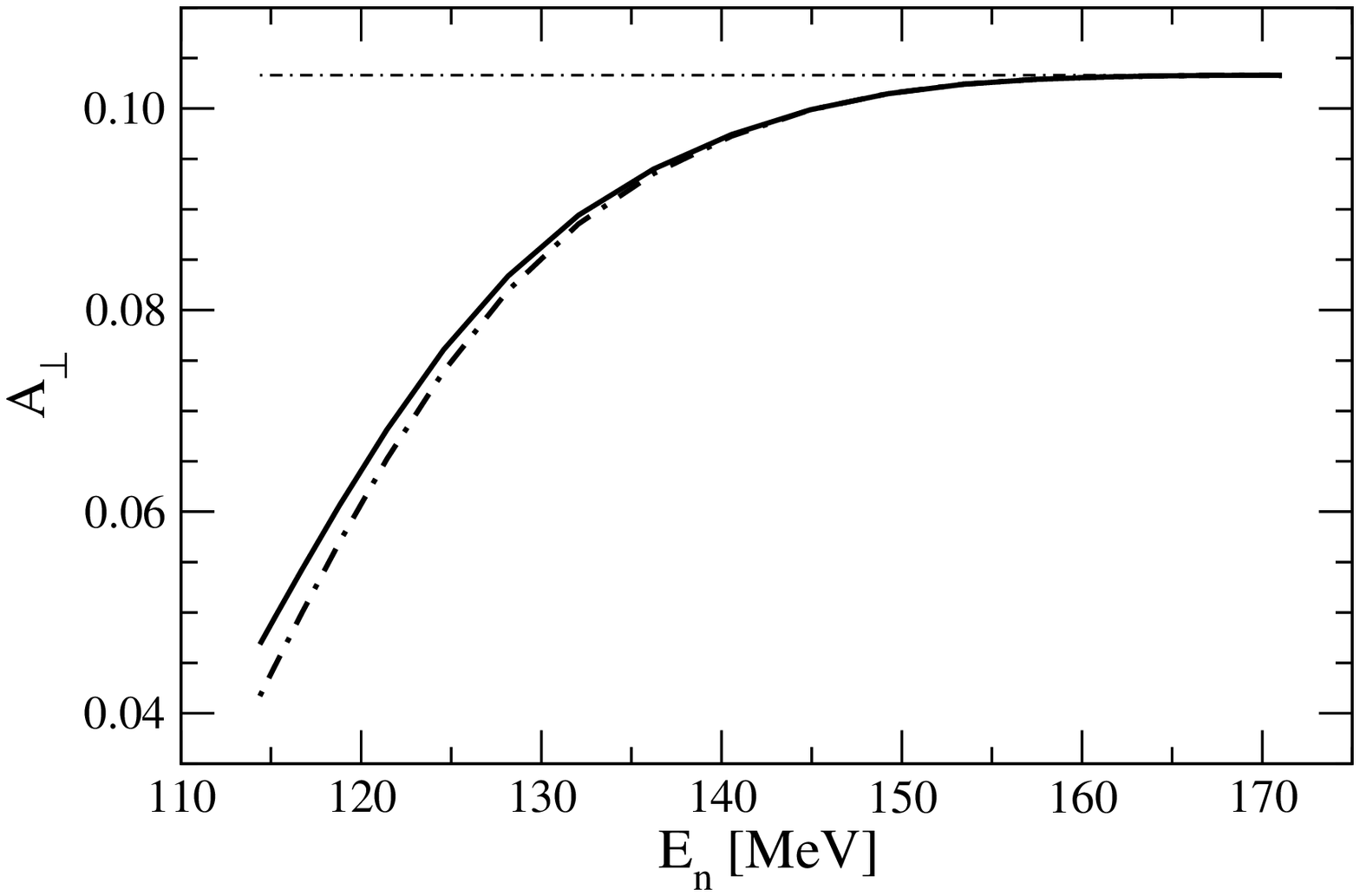,bb=5 350 550 750,clip=true,width=13cm}
\caption{\label{aperp.fsi23.npp}
The same as in Fig.~\ref{apara.fsi23.npp} for the perpendicular 
asymmetry $A_\perp$.
      }
\end{center}
\end{figure}

The situation for the neutron emission shown in Figs.~\ref{apara.fsi23.npp} 
and \ref{aperp.fsi23.npp} 
is much simpler and we do not observe so much sensitivity to different 
dynamical components.  The $t$-matrix is anyway forced to act in the 
total isospin $t=1$ states. Since additionally for the highest neutron energies 
(the lowest subsystem $23$ energies) the nucleon-nucleon interaction is restricted 
to $s$-waves, that implies that only the  $^1S_0$ partial wave 
should be important. This expectation is confirmed by our results. 


\begin{figure}[htb]
\begin{center}
\epsfig{file=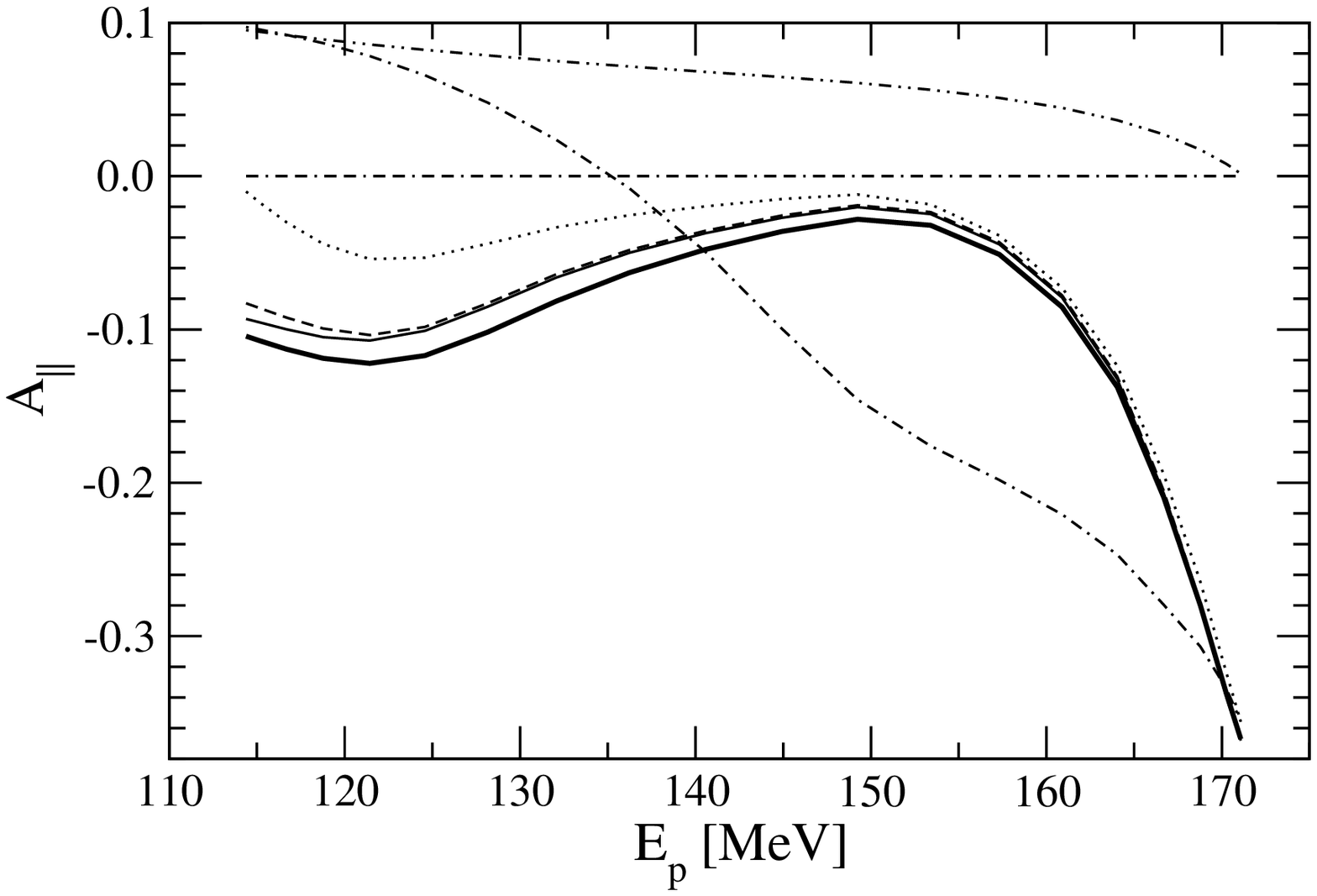,bb=5 350 550 750,clip=true,width=13cm}
\caption{\label{apara.full.ppn}
The parallel asymmetry $A_\parallel$ for the proton ejection in the
virtual photon direction as a function of the ejected proton energy $E_p$
for the electron configuration from Table~\ref{tab1} under
different dynamical treatments of FSI.
The double-dashed-dot line shows the PWIA prediction with the principal $S$-state
and the double-dotted-dash line the PWIA prediction with full $^3$He.
Further we show again the $FSI23$ predictions 
with the $^3$He restricted to the principal $S$-state
(dash-dotted line) and full $^3$He (dotted line). Results with the full inclusion 
of FSI and no MEC are plotted with the dashed line. The thin solid line 
represents the predictions which include the $\pi$- and $\rho$-like MEC 
and finally the thick solid line 
shows our best calculations involving in addition the UrbanaIX 3N force.
      }
\end{center}
\end{figure}

\begin{figure}[htb]
\begin{center}
\epsfig{file=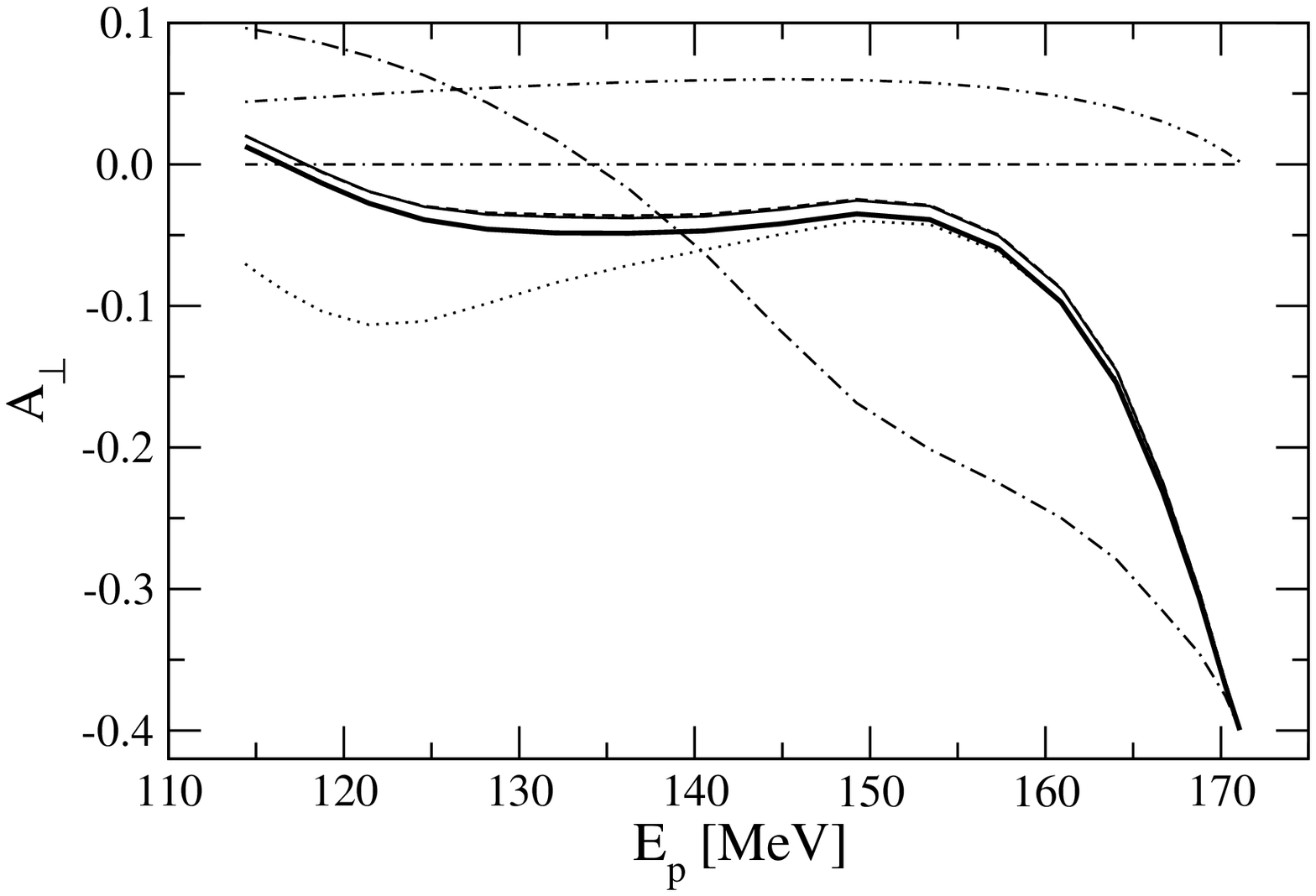,bb=5 350 550 750,clip=true,width=13cm}
\caption{\label{aperp.full.ppn}
The same as in Fig.~\ref{apara.full.ppn} for the perpendicular 
asymmetry $A_\perp$.
      }
\end{center}
\end{figure}

\begin{figure}[htb]
\begin{center}
\epsfig{file=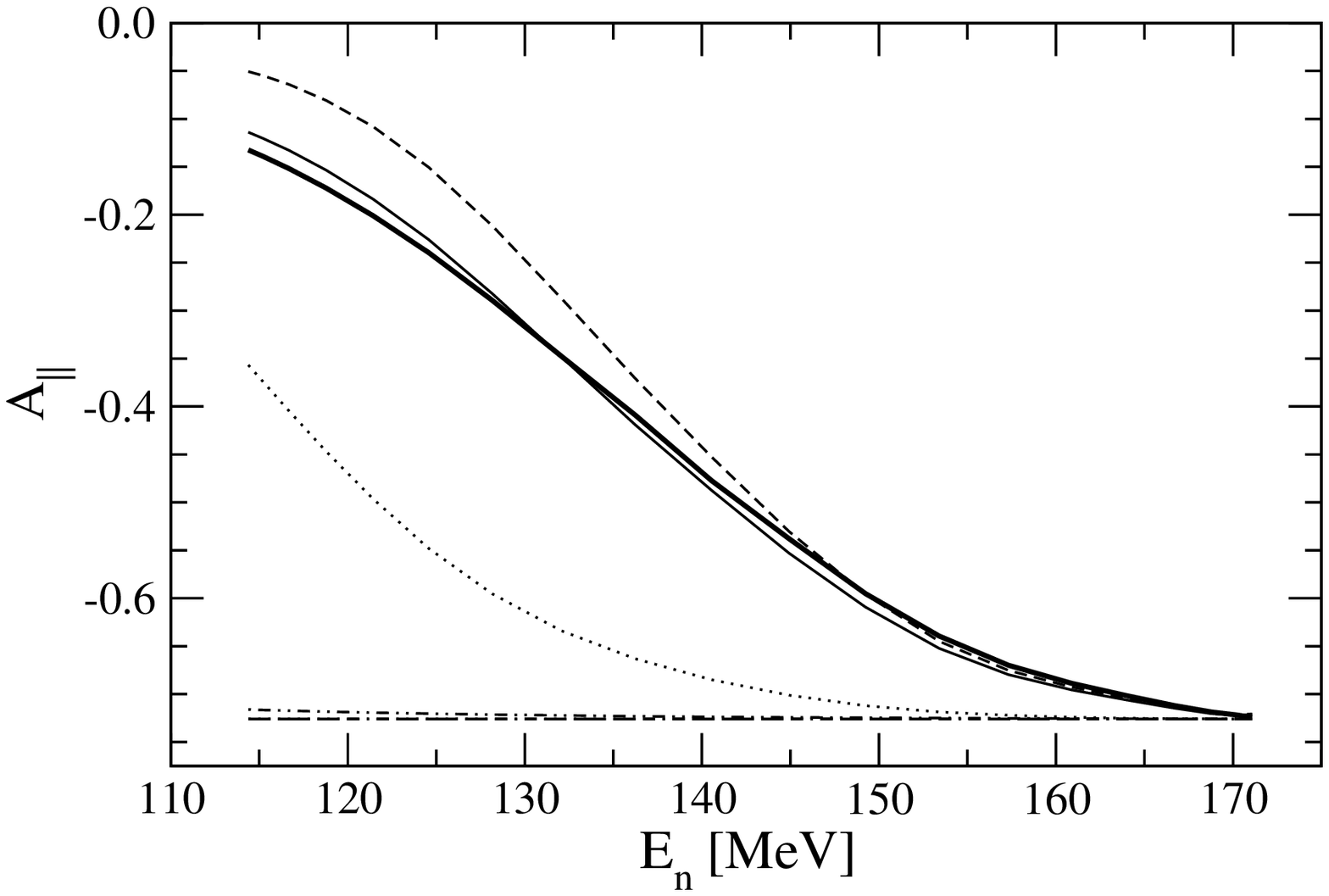,bb=5 350 550 750,clip=true,width=13cm}
\caption{\label{apara.full.npp}
The same as in Fig.~\ref{apara.full.ppn} for the neutron knock-out.
      }
\end{center}
\end{figure}

\begin{figure}[htb]
\begin{center}
\epsfig{file=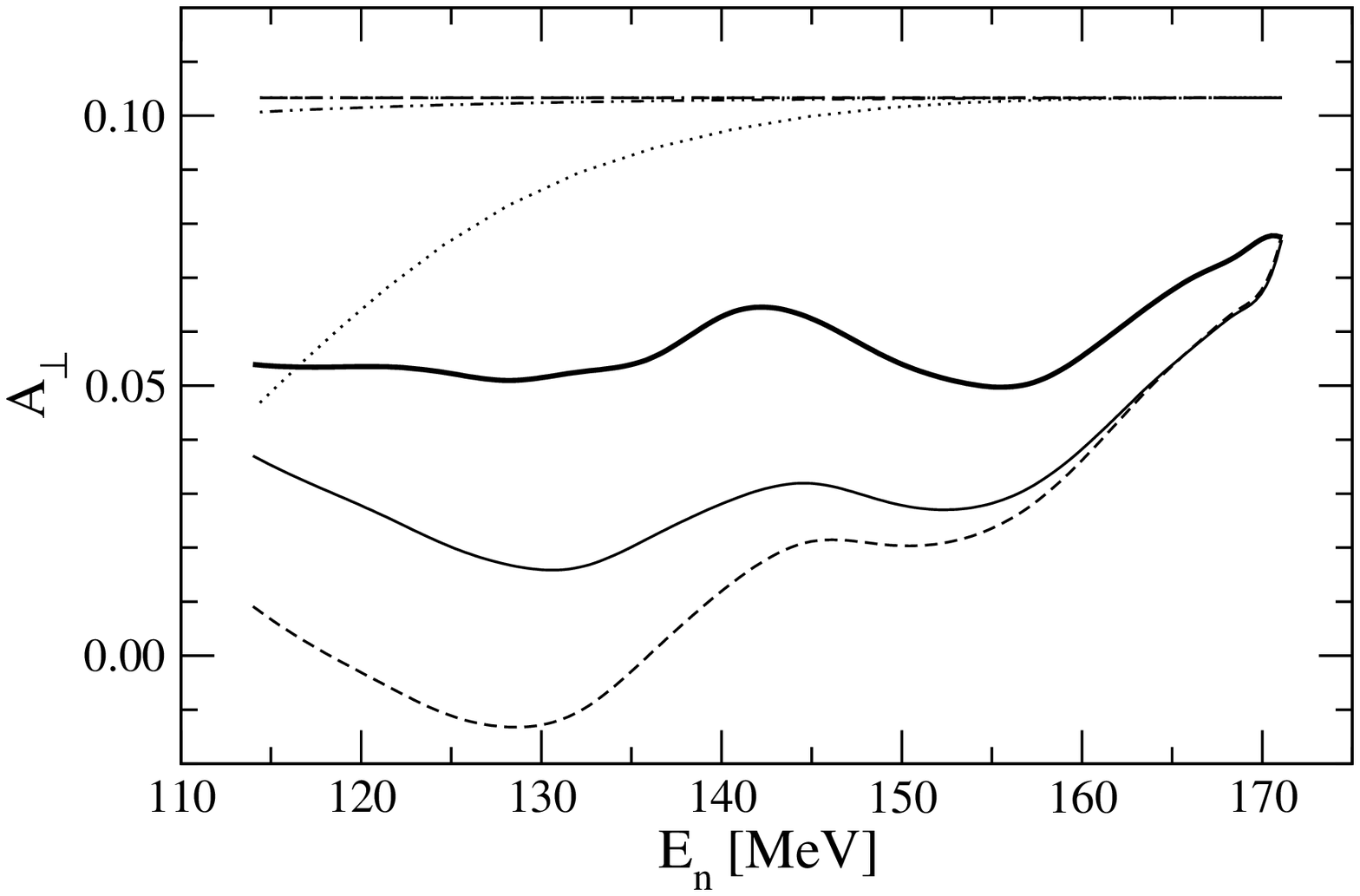,bb=5 350 550 750,clip=true,width=13cm}
\caption{\label{aperp.full.npp}
The same as in Fig.~\ref{apara.full.npp} for the perpendicular
asymmetry $A_\perp$.
      }
\end{center}
\end{figure}

In the group of figures \ref{apara.full.ppn}--\ref{aperp.full.npp}
we demonstrate results for much more complicated dynamical 
frameworks. We show first the results based on the full treatment of FSI.
Then we add to our single nucleon current the $\pi$- and $\rho$-like 
meson exchange currents. Finally we show the results where on top of all that
the UrbanaIX 3N force is present both for the initial $^3$He bound state 
and for the final scattering states.
For proton emission the $FSI23$ approximation but taking the full 
$^3$He state into consideration turns out to be satisfactory at the upper
end of the energy spectrum. This is valid for the both asymmetries.
In the case of neutron emission the situation is different and the full 
dynamics, especially for $A_\perp$ is required. 
It is only in the case of $A_\parallel$ that at the highest 
neutron energies all curves coincide. 

As pointed out before \cite{spindep,Groningen04}
that means that the extraction of $G_M^n$ from a measurement 
of the parallel asymmetry $A_\parallel$ seems to be quite model 
independent. This is not the case for the extraction 
of $G_E^n$ from a measurement of the perpendicular asymmetry 
$A_\perp$, which shows more sensitivity to different 
dynamical ingredients (see Fig~\ref{aperp.full.npp}).
To minimize the effects from complicated dynamics, measurements are 
performed on top of the quasi-elastic peak. 
Since the cross section drops very fast for the neutron energies below 
the maximal one (see Fig~\ref{sigma.full.npp}), $A_\perp$ receives 
main contributions from the regions where the model dependence 
is somewhat reduced.
  

\begin{figure}[htb]
\begin{center}
\epsfig{file=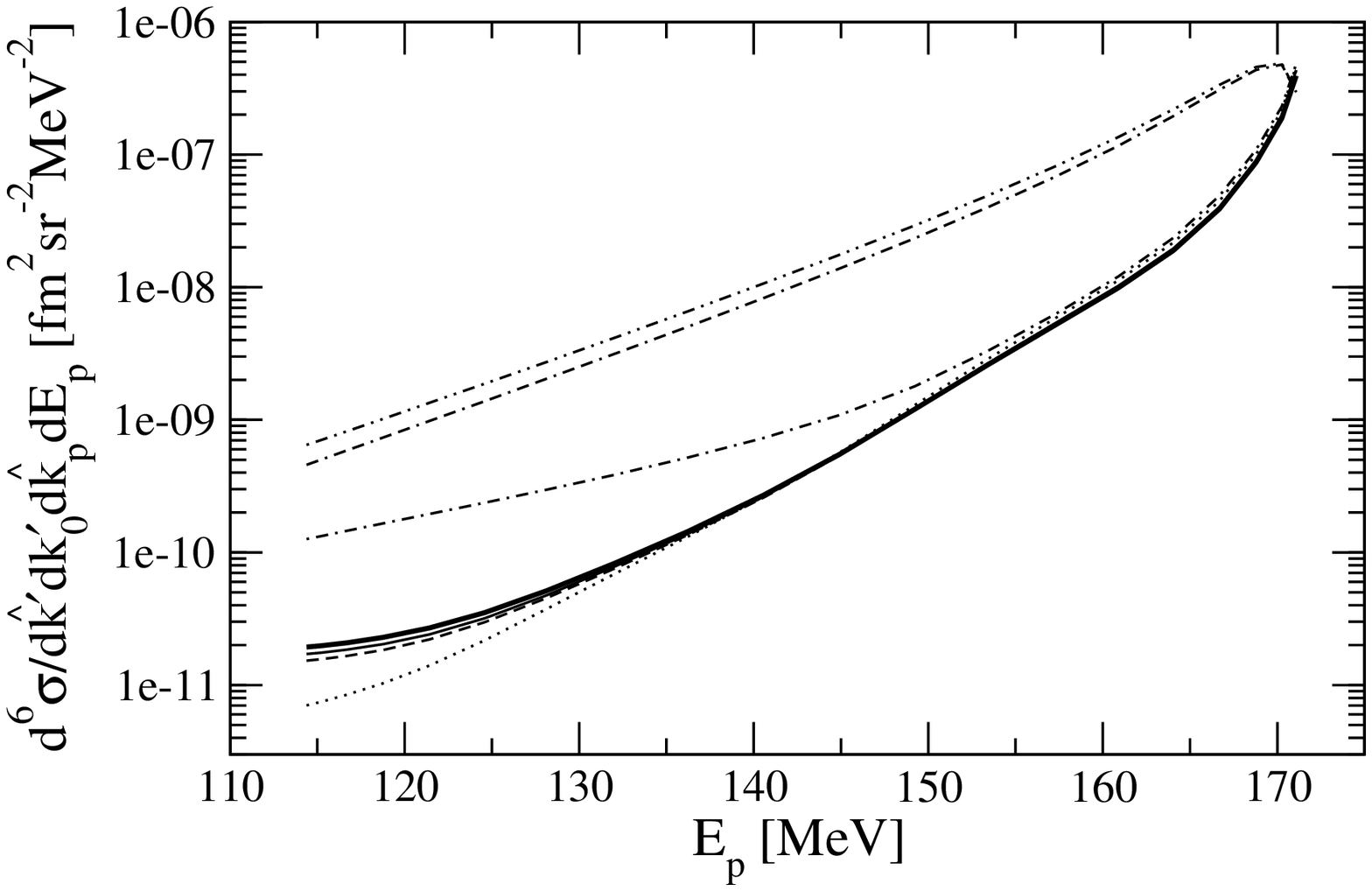,bb=5 350 550 750,clip=true,width=13cm}
\caption{\label{sigma.full.ppn}
The six fold differential cross section 
for the proton ejection in the
virtual photon direction as a function of the ejected proton energy $E_p$
for the electron configuration from Table~\ref{tab1} under
different dynamical treatments of FSI. Curves as in Fig.~\ref{apara.full.ppn}.
      }
\end{center}
\end{figure}

\begin{figure}[htb]
\begin{center}
\epsfig{file=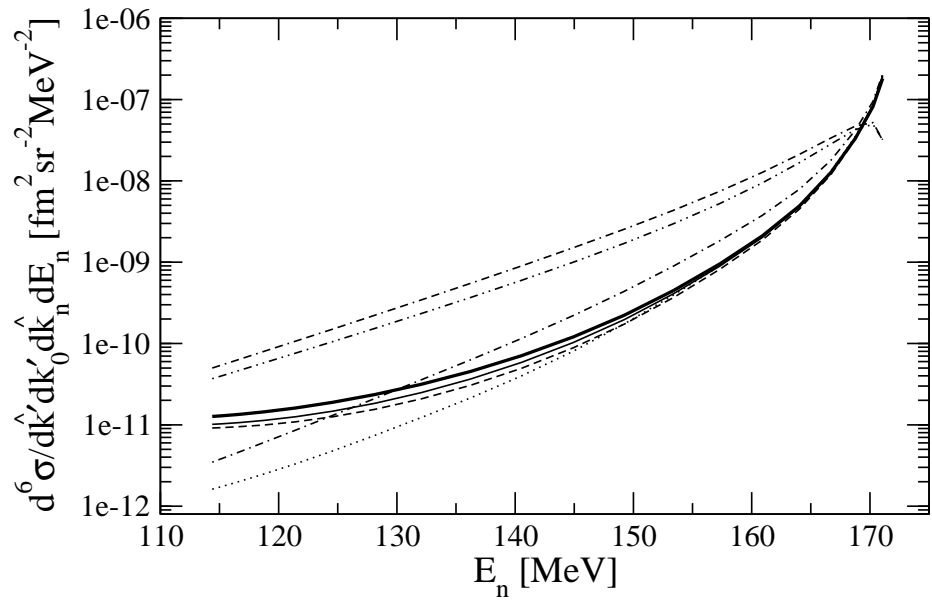,bb=5 350 550 750,clip=true,width=13cm}
\caption{\label{sigma.full.npp}
The same as in Fig.~\ref{sigma.full.ppn} for the neutron knock-out process.
      }
\end{center}
\end{figure}

Finally in Figs.~\ref{sigma.full.ppn} and \ref{sigma.full.npp} 
we show for the sake of completeness our predictions
for the six fold differential cross sections both for the proton and
neutron knock-out processes.

\clearpage

\section{Summary}
\label{sec:5}

The present paper is motivated by a recent experiment~\cite{Achenbach05},
where for the first time the  $A_\parallel$ and $A_\perp$ asymmetries 
were measured for proton emission in the two- and three-body breakup of $^3$He.
We present results for one of the electron kinematics measured 
in ~\cite{Achenbach05}.
For the 
$ \vec{^3{\rm He}} ({\vec e},e' p)d$
process this paper is a continuation of work in \cite{spindep}, where 
the spin dependent momentum distributions of proton-deuteron 
clusters in polarized $^3$He were investigated. Thus we can confirm that 
choosing the so-called parallel kinematics and small final deuteron momenta, 
information about proton polarization in $^3$He is directly available. 
We found that in such a case the polarizations extracted 
from the parallel and perpendicular asymmetries 
are not independent but simply related. This relation has been to some extent
confirmed in ~\cite{Achenbach05}.
For these specific kinematical conditions FSI (including the 3N force effects) 
and MEC do not play a big role and the PWIA picture is sufficient. 
One should exploit this opportunity and obtain all possible information 
about $^3$He. On the other hand, this could also be a method to measure the proton 
electromagnetic form factors, even though they are 
known from direct electron scattering on a proton target.
Such a measurement on $^3$He would verify our knowledge about this nucleus
and help set a limit on medium corrections of the form factors. 

The situation for the 
$ \vec{^3{\rm He}} ({\vec e},e' p)pn$
reaction is more complicated since the simplest PWIA approximation
is not valid. For the proton emission we find a lot of sensitivity to 
the smaller $^3$He wave function components because for the main 
principal $S$-state of $^3$He the asymmetries are zero.
FSI has to be taken into account but for the parallel
kinematics and high emitted proton energies it can be approximated 
by a simpler $FSI23$ prescription. This is in agreement with the 
results of a study performed in \cite{spectral}.
We find, however, that no picture of electron scattering 
on a polarized proton arises. The reason is that even at the highest 
proton energies partial waves with spin $s=0$ and $s=1$ contribute.

For the 
$ \vec{^3{\rm He}} ({\vec e},e' n)pp$
reaction we see again (see \cite{sensitivity}) different sensitivities 
of the $A_\parallel$ and $A_\perp$ asymmetries
to the dynamical ingredients of our Faddeev framework.
This proves that the extraction of $G_M^n$ from a measurement
of the parallel asymmetry $A_\parallel$ would be very simple.
This is not quite the case for the extraction
of $G_E^n$ from a measurement of the perpendicular asymmetry
$A_\perp$, where corrections from FSI, MEC and 3N forces 
would play a more important role. The theoretical uncertainties 
can be, however, minimized by a proper choice of experimental 
conditions.

Finally, we would like to emphasize that the results reflect 
our present day understanding of the reaction mechanism 
and the structure of $^3$He. 
Therefore new data for the processes addressed
in this paper would be extremely useful.

\acknowledgments
This work was supported by the Polish Committee for Scientific Research
under grant no. 2P03B00825, by the NATO grant no. PST.CLG.978943,
and by DOE under grants nos. DE-FG03-00ER41132 and DE-FC02-01ER41187.
One of us (W.G.) would like to thank the Foundation for Polish Science
for the financial support during his stay in Krak\'ow.
We would like to thank Dr. Rohe and Dr. Sirca for reading 
the manuscript and important remarks. 
The numerical calculations have been performed on the Cray SV1 and
on the IBM Regatta p690+ of the NIC in J\"ulich, Germany.



\begin{thebibliography}{99}

\bibitem{Fentometer} J. L. Forest, V. R. Pandharipande, Steven C. Pieper,
                     R. B. Wiringa, R. Schiavilla, A. Arriaga, 
                     Phys. Rev. C{\bf 54}, 646 (1996).

\bibitem{spindep} J. Golak, W. Gl\"ockle, H. Kamada, H. Wita{\l}a, 
                  R. Skibi\'nski, A. Nogga,
                  Phys. Rev. C{\bf 65}, 064004 (2002).

\bibitem{Achenbach05} P. Achenbach {\em et al.}, nucl-ex/0505012.

\bibitem{Gloecklebook} W. Gl\"ockle, {\em The Quantum Mechanical
                       Few-Body Problem} (Springer-Verlag, Berlin, 1983).

\bibitem{Donnelly} T. W. Donnelly, A. S. Raskin, Ann. Phys. (N.Y.) {\bf 169}, 247 (1986).

\bibitem{report} J.~Golak, R.~Skibi\'nski, H.~Wita{\l}a, W.~Gl\"ockle, 
                 A.~Nogga, H. Kamada, Phys. Rep. {\bf 415}, 89 (2005).

\bibitem{AV18} R. B. Wiringa, V.G.J. Stoks, R. Schiavilla,
               Phys. Rev. C{\bf 51}, 38 (1995).

\bibitem{UrbanaIX} B. S. Pudliner, V. R. Pandharipande,
                   J. Carlson, Steven C. Pieper and R. B. Wiringa,
                   Phys. Rev. C{\bf 56}, 1720 (1997).

\bibitem{Riska85} D. O. Riska, Phys. Scr. {\bf 31}, 107 (1985);
                  D. O. Riska, Phys. Scr. {\bf 31}, 471 (1985).

\bibitem{Groningen04} J. Golak, R. Skibi\'nski, H. Wita{\l}a, W. Gl\"ockle, 
                      A. Nogga, H. Kamada, 
                      in: N. Kalantar-Nayestenaki, R.G.E. Timmermans, B.L.G. Bakker (Eds.),
                      Few-Body Problems in Physics, AIP Conference Proceedings No. 768,
                      AIP, Melville, NY, 2005, p. 91.

\bibitem{spectral} J. Golak, H. Wita{\l}a, R. Skibi\'nski, W. Gl\"ockle, A. Nogga, H. Kamada,
                   Phys. Rev. C{\bf 70}, 034005 (2004).

\bibitem{sensitivity} J. Golak, W. Gl\"ockle, H. Kamada, H. Wita{\l}a, R. Skibi\'nski, A. Nogga,
                      Phys. Rev. C{\bf 65}, 044002 (2002).


%
%
%

\end{thebibliography}
\end{document}